\documentstyle[aps,latexsym,epsfig,graphicx,amssymb]{revtex}

\newcommand{\bu}{{\mathbf u}}

\newcommand{\bx}{{\mathbf x}}

\newcommand {\low}[1] {{\mbox{\scriptsize #1}}}

\begin{document}

\title{Mixing and reaction efficiency in closed domains}

\author{S. Berti$^{1}$, D. Vergni$^{2}$, 
        F. Visconti$^{3}$ and A. Vulpiani $^{3,4}$}
\address{$^1$ Dipartimento di Fisica Generale, Universit\`a di Torino,
Via Pietro Giuria 1, I-10125 Torino, Italy} 
\address{$^2$ Istituto Applicazioni del Calcolo (IAC) - CNR,
Viale del Policlinico, 137 I-00161 Roma, Italy}
\address{$^3$ Dipartimento di Fisica - Universit\`a ``La Sapienza'',
Piazzale Aldo Moro 2, I-00185 Roma, Italy}
\address{$^4$ Istituto Nazionale di Fisica della Materia (UdR),
SMC (Roma 1), Istituto Nazionale di Fisica Nucleare (Roma 1),\\ 
Piazzale Aldo Moro 2, I-00185 Roma, Italy}
\maketitle

\begin{abstract}
We present a numerical study of mixing and reaction efficiency in
closed domains.  In particular we focus our attention on laminar flows.  
In the case of inert transport the mixing properties of the flows
strongly depend on the details of the Lagrangian transport.  
We also study the reaction efficiency. Starting with a little spot of
product we compute the time needed to complete the reaction in the
container.  We found that the reaction efficiency is not strictly
related to the mixing properties of the flow. 
In particular, reaction acts as a ``dynamical regulator''.
\end{abstract}
\vspace{0.5truecm}

\noindent{Keyword: Laminar Reacting Flows; non Asymptotic Propagation}\\
{PACS numbers: 05.45.-a; 47.70.Fw}
\section{Introduction}
\label{sec:1}
Transport of reacting species advected by laminar or turbulent flows
(advection reaction diffusion -- ARD -- systems), is an issue of obvious
interest in many fields, e.g. population dynamics (propagation of
plankton in oceanic currents~\cite{abr98}), reacting chemicals in the
atmosphere (e.g., ozone dynamics~\cite{polvortex}), complex chemical
reactions and combustion~\cite{zel38}. For recent interesting
experimental studies see Ref.~\cite{expe}.

The simplest non trivial case of ARD is described by a scalar field 
$\theta(\bx, t)$, which represents the concentration of reaction 
products, such that $\theta$ is equal $1$ in the regions where the 
reaction is over (the stable phase), and $\theta$ is zero where  
fresh material is present (the unstable phase). The field
$\theta(\bx, t)$ evolves according to the following equation:
\begin{equation}
   \partial_t \theta + (\bu \cdot \nabla) \theta = D_\low{0} \Delta \theta + 
   {1 \over \tau} f(\theta),
   \label{eq:ARD}
\end{equation}
where $\bu$ is a given incompressible velocity field and $D_\low{0}$ is the 
molecular diffusion coefficient. Of course the reaction is described 
by the term $f(\theta)/\tau$ where $\tau$ is the time scale of 
the chemistry.

The form of the reacting term $f(\theta)$ depends on the problem under 
investigation; a rather popular case is the so-called Fisher-Kolmogorov-
Petrovsky-Piskunov (FKPP) nonlinearity~\cite{fkpp}
$f(\theta) = \theta (1-\theta)$, 
which describes the auto-catalytic process $A + B \rightarrow 2A$ 
(in such a case $\theta$ is the concentration of the species $A$).
This nonlinearity belongs to the more general class of FKPP-type
non-linearity characterized by having the maximum slope of $f(\theta)$ 
in $\theta\!=\!0$. Those non-linearity terms give rise to the 
so called {\em pulled} fronts for which front dynamics can be
understood by linear analysis, since it is essentially determined by
the $\theta(x,t)\approx 0$ region (the front is pulled by its
leading edge). In the case of front propagation in reaction-diffusion 
systems (i.e., with $\bu=0$) it is possible to show~\cite{mur93} that, 
for FKPP-like non linearity, a moving front (i.e., an ``invasion'' 
of the stable phase, $\theta = 1$, in the unstable one, $\theta = 0$) 
develops  with propagation speed given by
$v_\low{0} = 2\sqrt{D_\low{0} f'(0)/\tau }$.

Another important class of non-linearity terms is the non-FKPP-type, for
which the maximal growth rate is not realized at $\theta\!=\!0$ but at
some finite value of $\theta$, where the details of the nonlinearity
of $f(\theta)$ are important.  In this case front dynamics is often
referred to as {\it pushed}, meaning that the front is pushed by its
(nonlinear) interior. At variance with the previous case,
the determination of the front speed now requires a detailed 
non linear analysis.
It is not possible to give a general result for
the front speed, but only the bound (when $\bu = 0$)
$2\sqrt{D_\low{0}f'(0)} \leq v_\low{0} < 
 2\sqrt{D_\low{0}\sup_\theta \left\{{f(\theta) \over
\theta}\right\}}\,$~\cite{mur93} can be obtained.
An important example of non-FKPP-type
non-linearity is given by the so called Arrhenius term: 
$f(\theta) = (1-\theta) e^{-\theta_c/\theta}$.

In the following we principally adopt the FKPP-type nonlinearity.
However, in order to investigate the relevance of $f(\theta)$,
we also discuss the non FKPP-type nonlinearity.

If we suppress the reacting term $f(\theta)/\tau$ 
in (\ref{eq:ARD}) we obtain the advection-diffusion equation which rules 
the evolution of the concentration $P(\bx, t)$ of inert particles
\begin{equation}
   \partial_t P + (\bu \cdot \nabla) P = D_\low{0} \Delta P\,.
   \label{eq:AD}
\end{equation}
Let us underline that for both the processes (\ref{eq:ARD}) and
(\ref{eq:AD}) (reactive and inert transport, respectively), one can
face different classes of problems, namely the asymptotic and non
asymptotic ones~\cite{bof00}.

By asymptotic properties we mean the features of Eqs.
(\ref{eq:ARD}) and (\ref{eq:AD}) at long times and large spatial
scales, i.e., much larger than the typical length $\ell$ of $\bu$.  
In such a limit, under rather general conditions~\cite{may99},
Eq.~(\ref{eq:AD}) reduces to an effective diffusion equation
\begin{equation}
   \partial_t P  = \sum_{i,j}D_{ij}^e \partial^2_{ij} P
   \label{eq:Deff}
\end{equation}
where the effective diffusion tensor $D_{ij}^e$ depends, often in a 
nontrivial way, on $\bu$. In a similar way for the ARD problem if we 
start with a localized region in which $\theta = 1$ (elsewhere $\theta = 0$), 
one asymptotically has a front propagation with a front speed 
$v_\low{f}$ depending on $\tau$, $D_\low{0}$ and $\bu$~\cite{con00}.

Although the asymptotic problems are well defined from a mathematical 
viewpoint, sometimes their relevance in real life is rather poor. 
Often, e.g., in geophysics, the spatial size, $L$, of the system
is comparable with the typical length of the flow, $\ell$, so 
it is not possible to use Eq. (\ref{eq:Deff}) for the dispersion of 
passive inert scalar fields and it is necessary to treat the problem 
using some indicators able to go beyond the diffusion coefficient. 
Such a problem had been studied, for example, in Refs.~\cite{bof00,bof01}
using the Finite Size Lyapunov Exponents which properly
characterize the transport mechanism at a given (spatial) scale.
\\
In a similar way, considering cases with $L$ not too large
compared with $\ell$, one has nontrivial features also in ARD systems.
As an example we can mention Ref.~\cite{lop02}
where it had been found that the burning efficiency in a closed domain
does not increase for large values of the strength of the velocity
field.

As a general remark, we stress that both in inertial 
and reactive transport the Eulerian turbulence has a minor role. 
As examples we can mention the Lagrangian chaos, i.e., the irregular 
behavior of passive tracers also for laminar flow~\cite{Aref}
and the poor role of the presence of small scales in the 
velocity field for front propagation~\cite{Cencini2003}.

In this paper we discuss the mixing and reaction efficiency 
[in Eqs.~(\ref{eq:AD}) and~(\ref{eq:ARD}), respectively]
in systems advected by a given velocity field, $\bu$, in closed
domains. For mixing efficiency we intend the capacity of a flow 
to spread particulate inert material starting from a small region
over the whole system domain. 
When the material is chemically active, the interesting question
concerns the times needed to complete the reaction in the system
domain, that we call reaction efficiency. 
At an intuitive level one could expect a link between mixing 
and reaction efficiency, because both are related to the transport 
properties of the velocity field, but we show there exist cases 
in which this relation is very weak.

We will see that for the inert transport problem
(\ref{eq:AD}), the mixing efficiency strongly depends on the features
of the dynamical system
\begin{equation}
   {{\mathrm d}\bx \over {\mathrm d}t}  = \bu\,\,,
   \label{eq:DS}  
\end{equation}
in particular if large scale chaos is present or not. 
On the contrary for the reacting case (\ref{eq:ARD}), the presence of 
large-scale chaos has a minor role.
This result is rather close to those obtained in other 
subtle issues such as the classical limit of quantum 
mechanics~\cite{quantum}, or metastable balance between
chaos and diffusion~\cite{Patta}.
\\
The paper is organized as follows:
In Section II we introduce two flow models for the velocity field {\bf u}
and we discuss the mixing efficiency (for inert particles) in closed
domains at varying the chaotic properties of Eq.~(\ref{eq:DS}).
Section III is devoted to the burning efficiency in the reactive case. 
We will see that, at variance with the inert case, the details of the
Lagrangian transport are not very relevant.
In Sec. IV the reader finds remarks and conclusions.
\section{Mixing Efficiency of inert transport}
\label{sec:2}
The limit case $D_\low{0} = 0$ in Eq.~(\ref{eq:AD})
is related to the Lagrangian deterministic motion (\ref{eq:DS}).
When molecular diffusion is present, we have to consider the Langevin
equation obtained by adding a noisy term to (\ref{eq:DS})
\begin{equation}
   {{\mathrm d}\bx \over {\mathrm d}t} = 
	\bu + \sqrt{2D_\low{0}} \mbox{\boldmath $\eta$}\,\,,
   \label{eq:La}
\end{equation}
where $\eta$ is a white noise. Therefore, Eq.~(\ref{eq:AD}) is 
nothing but the Fokker-Planck equation related to (\ref{eq:La}).
\\
Since we are interested in the mixing in closed domains $\Omega$,
we have to specify the boundary conditions: in the case $D_\low{0} = 0$
the perpendicular component of $\bu$ on the border $\partial \Omega$ 
must be zero.
In $2D$ it is very easy to impose this constraint. Writing 
$\bu = (\partial_y \psi, - \partial_x \psi)$, one has 
$\psi = {\mathrm const}$ for $\bx$ on $\partial \Omega$. 
Analogous in the case $D_\low{0} > 0$ is the no-flux condition
$\frac{\partial P}{\partial x_{\bot}}|_{\partial \Omega} = 0$. 
In terms of the Langevin equation (\ref{eq:La}) this corresponds 
to a reflection of the trajectory $\bx(t)$ on $\partial \Omega$.

In the following we will limit our attention to $2D$ cases, i.e., 
\begin{equation}
   \psi(x,y,t)  = \psi_\low{0}(x,y) + \epsilon \psi_1(x,y,t)\,\,,
   \label{eq:psi-t}
\end{equation}
where $\psi_1$ is time-periodic function of period $T$. 

First we analyze the case of $D_\low{0} = 0$. 
If $\epsilon = 0$, Eq. (\ref{eq:DS}) cannot exhibit
a chaotic behavior. On the other hand, if $\epsilon \neq 0$ one can
have chaos (and this is the typical feature) around the separatrix
(periodic orbit of infinite period). At small $\epsilon$ chaos is
restricted to a limited region and it has just a poor role for the mixing in
$\Omega$. In order to have ``large-scale'' chaotic mixing (i.e., the
possibility to cross the unperturbed separatrix) $\epsilon$ must be
larger than a certain critical value $\epsilon_c$ (which depends on
$T$). This is the essence of the celebrated ``overlap of the resonances 
criterion'' by Chirikov~\cite{chi79,lic93}.
\\
It is easy to realize that, if $D_\low{0} \neq 0$, after a
sufficiently long time, tracers will invade the whole basin, i.e.,
there will be no more barriers to transport. The interesting question
in such a case is to understand the mechanism which determines the 
mixing time, i.e., the time to have a spatial homogenization
due to the mixing process.
\\
Consider as initial condition a distribution $P(\bx,0)$ localized
around $\bx_0$: in Lagrangian terms an ensemble of particles initially
concentrated in a small region of size $\delta_\low{0}$. 
There are two limit cases in which it is simple to understand
the local transport properties, namely for very large scale $r$, 
i.e., $r \gg \ell$, and for very small scales, i.e., $r \ll \ell$. 
Let us remind that $\ell$ is the typical spatial scale of the flow $\bu$.
In the last case, if $D_\low{0} = 0$ and the dynamical system given 
by Eq.~(\ref{eq:DS}) is chaotic, we have, 
if $|\delta\bx(t)| \ll \ell$
\begin{equation}
  |\delta\bx(t)| \simeq |\delta\bx(0)|e^{\lambda t}\,.
  \label{eq:lyap}
\end{equation}
From the previous equation one could naively conclude that
in a closed domain of size $L$ the typical mixing time is
\begin{equation}
  \tau_m \sim {1 \over \lambda} \log {L\over \delta_\low{0}} 
         \sim {1 \over \lambda}\,.
  \label{eq:timelyap}
\end{equation}
Of course this is a very crude conclusion which does not consider
some basic facts~\cite{bof00,bof01}:
\begin{description}
\item [(i)] the $\lambda$ in Eq.~(\ref{eq:timelyap}) usually depends on the
  initial condition, so instead of $\lambda$ one could consider
  the Kolmogorov-Sinai entropy~\cite{lic93}
  \begin{equation}
    h_\low{KS} = \int_\Omega \lambda(\bx) {\mathrm d}\mu(\bx)
    \label{eq:hKS}
  \end{equation}
  where $\lambda(\bx) = \lim_{t \to \infty} 
                              \lim_{\delta_\low{0} \to 0}
		{1\over t} \ln {|\delta \bx(t)| \over \delta_\low{0}}$
  with the initial condition starting from $\bx$. Equation~(\ref{eq:hKS})
  follows from the symplectic nature of our bidimensional problem;
  therefore $\lambda(\bx)$ can be positive or zero~\cite{lic93};
\item [(ii)] the existence of barriers (in the non overlap cases);
\item [(iii)] the effect of noise (i.e., molecular diffusivity);
\item [(iv)] in Eq.~(\ref{eq:timelyap}) one assumes the possibility 
  to linearize the equation for $\delta \bx(t)$.
\end{description}
In the opposite limit $r \gg \ell$ the asymptotic transport 
is described by the effective Fick equation~(\ref{eq:Deff}), 
and the mixing time is simply
\begin{equation}
  \tau_m \sim {L^2 \over D^e}\,,
  \label{eq:timefick}
\end{equation}
where $D^e$ is the effective diffusion coefficient.
For the sake of simplicity we ignore the tensorial nature
of $D_{ij}^e$.
Also in this case there are some caveats:
\begin{description}
\item [(i)] Equation~(\ref{eq:timefick}) holds only if $L\gg \ell$;
\item [(ii)] Equation~(\ref{eq:timefick}) ignores (possible important) transient effects.
\end{description}
In this paper we treat the non asymptotic case, i.e., $L \sim \ell$.

Let us discuss a rather natural procedure for the characterization
of the mixing efficiency. Introduce a coarse graining of the phase
space $\Omega$ [note that in this case the phase space coincides with
the physical space $(x,y)$], with $N$ square cells of size $\Delta$.
As initial condition we take ${\mathcal N} \gg 1$ particles 
in a unique cell. At time $t > 0$ we compute the quantity
\begin{equation}
   P_i(t) = {n_i(t) \over {\mathcal N}}\,,
   \label{eq:perpar}
\end{equation}
which gives the percentage of particles in the $i$-th cell 
[$n_i(t)$ being the number of particles in the $i$-th cell at time $t$].
Then, we define the ``occupied area'', $A(t)$, as
the percentage of ``occupied'' cells. With the term ``occupied'' we mean
the number of particles in the cell is larger than a preassigned
quantity (e.g. $25\%$) of the average number of particles per cell
in the uniform dispersion situation, i.e. $P_i(t) > c / N$
where $c=0.25$
\begin{equation}
   A(t) = {1\over N} \sum_{i=1}^{N}
          \theta \left (P_i(t) - {c \over N} \right )\,\,
   \label{eq:occar}
\end{equation}
where $\theta(\cdot)$ is the step function.\\
As an indicator of the mixing efficiency, we compute the
\emph{mixing time} as the first time at which 
a given percentage, $\alpha$, of the total area is filled up 
\begin{equation}
   t_{\alpha} = \mbox{min} \{t\,:\:A(t)=\alpha\,\}.
   \label{eq:talpha}
\end{equation}

Another possible indicator of the mixing efficiency is the following:
\begin{equation}
   Q(t) = {1 \over N} \sum_{i=1}^{N} 
                          \left (P_i(t) - {1 \over N} \right )^2,
   \label{eq:var}
\end{equation}
which measures the average distance between the percentage
of particles in the cells at time $t$ and the percentage 
of particles referred to a uniform distribution. 
The system is perfectly mixed when $Q = 0$.
It is reasonable to expect that $Q(t)$ decreases exponentially
(at least at large times).
The behavior at large $t$ of the quantity $Q(t)$ is clearly related
to the spectrum of the operator
$${\mathcal L} = - (\bu \cdot \nabla) + D_\low{0} \Delta \,.$$
The largest eigenvalue $\gamma_\low{1}=0$ is in correspondence
with the eigensolution $\theta = $constant; if the second
eigenvalue $\gamma_\low{2} \neq 0$, then, at large times,
one has:
$$   Q(t) \sim e^{-2|\gamma_\low{2}| t}\,\,.$$
Of course $\gamma_\low{2}$ can depend both on the details of 
$\bu$ and the value of $D_\low{0}$~\cite{cer03}.
\subsection{Flow models}
\label{subsec:2.A}
Let us now introduce the velocity fields we considered, namely
the Meandering jet and the Stokes flow.
\subsubsection{Meandering jet}
\label{subsec:2.A.1}
The Meandering jet flow \cite{bow91,cen99}, first introduced as a
kinematic model for the Gulf Stream, is often used to describe western
boundary current extensions in the ocean. This flow has a periodic
spatial structure of wavelength $\ell$ (along the $x$-axis), 
characterized by the simultaneous
presence of regions with different dynamical properties: the jet core,
where the motion is ballistic, some recirculation zones where
particles tend to be trapped, and an essentially quiescent far
field. In a frame moving eastward with a velocity coinciding with the
phase speed, and after a proper nondimensionalization, its
stream function is
\begin{equation}
   \psi(x,y,t)  = -\tanh \left[\frac{y-B(t)\cos kx}
                                    {\sqrt{1+k^2B(t)^2\sin^2kx}}\right]+cy\,,
   \qquad\mbox{where}\qquad B(t)=B_0+\epsilon\cos(\omega t+\phi).
   \label{eq:mjet-sf}
\end{equation}

\begin{figure}[htbp]
\begin{center}
\epsfig{file=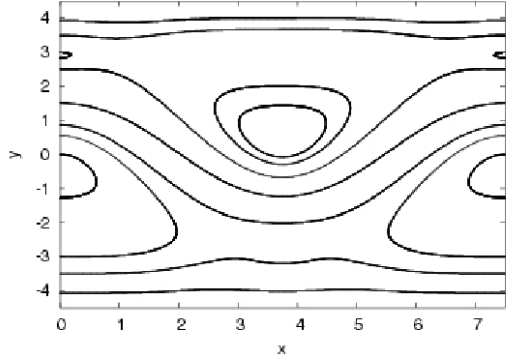,height=5cm,clip=}
\caption{Stationary $(\epsilon=0)$ meandering jet streamlines.}
\label{fig:mjet-sl}
\end{center}
\end{figure}

Here and in the following, we use the parameter values:
$k=2\pi/\ell=4\pi/15$, $B_0=1.2$, $\phi=\pi/2$, $c=0.12$, for the wave
number, the unperturbed meanders' amplitude, a perturbation's phase,
and the intensity of the far field, respectively; a sketch of the
streamlines for the stationary flow is presented in 
Fig.~\ref{fig:mjet-sl}. With these parameters, no particles
reach the far field, and no trajectories attain values in $|y|$ larger
than $4$ (even though, in general, we expect a low but nonzero
fraction of them to visit that area).

\begin{figure}[htbp]
\begin{center}
\vspace{1.0cm}
\epsfig{file=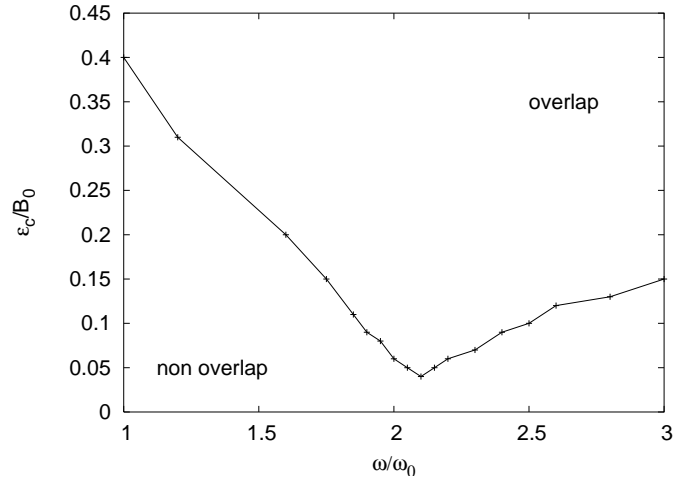,height=6.5cm,clip=}
\caption{Overlap of the resonances: $\epsilon_c$ vs $\omega$.
Here $\omega_0=0.25$ is the typical pulsation
of recirculation.}
\label{fig:overlap}
\end{center}
\end{figure}

The time dependence of the stream function is
sufficient to produce Lagrangian chaos. The ``chaoticity degree'' is
controlled by the two parameters $\epsilon$ and $\omega$.
Specifically, there exists a threshold value $\epsilon_c(\omega)$
determining a transition from \emph{local} to \emph{large-scale}
chaos, in agreement with Chirikov's overlap of the resonances
criterion~\cite{chi79,lic93}. In the first case $(\epsilon < \epsilon_c)$
chaos is confined to the stochastic layers around the separatrices,
while in the second one $(\epsilon > \epsilon_c)$ 
a large part of the
phase space is visited by a chaotic trajectory, indicating the
disappearance of any dynamical barriers to cross-stream transport
\cite{bof01}.  In the plane $(\omega,\epsilon)$ the relation
$\epsilon_c=\epsilon_c(\omega)$ defines a curve
(Fig.~\ref{fig:overlap}) which separates regions with different
dynamical properties, allowing one to discriminate between a \emph{nonoverlap} 
(local chaos, i.e., chaos in a bounded region of $\Omega$) 
and an \emph{overlap} (large scale-chaos) regime.
In the following we will discuss transport in a non asymptotic
situation, forcing the system to be in a small and closed basin, 
i.e., $0\leq x \leq 2\ell$.
We use periodic boundary conditions along
the $x$ axis; along the $y$ axis we set rigid boundary conditions in
order to maintain trajectories in the strip $|y|<4$ even in the presence
of a nonzero molecular diffusivity: particles reaching the
horizontal lines $y=-4,4$ are reflected backward.
We note here that, from a dynamical point of view, this system resembles
the stratospheric polar vortex \cite{polvortex}, once closed on itself in
a circular geometry. This vortex models a current of isolated air in the
high atmosphere, centered on the poles of the Earth, which has quite
an important role in the dynamics of stratospheric ozone.
\subsubsection{Stokes flow}
\label{subsec:2.A.2}
This is a simple model of cellular flow, often used in the past
because of its versatility, either from the experimental point of view
or from the computational one~\cite{vikh02,ot89}.
\\
Its spatial structure in the stationary case is rather simple: there are
only recirculation regions.  Once periodic time dependence is switched
on, stretching and folding of the streamlines can be seen and classic
``coffee and cream'' pathways take place.  The
flux is essentially driven from an upper ($V_{top}$) and a lower
($V_{bot}$) velocity; we choose to insert here time dependence 
in order to have Lagrangian chaos, setting $V_{top}=\cos{[\phi(t)]}$,
$V_{bot}=\sin{[\phi(t)]}$.
The stream function is
\begin{equation}
   \psi(x,y,t) = {1 \over 2} \left \{ 
 (y + 1) \cos[\phi(t)] + (y-1)\sin[\phi(t)] \right \}
  (1-x^2)(1-y^2)
   \label{eq:stokes}
\end{equation}
where $\phi(t) = 2\pi t/T$, and $T$ is the control parameter.
At varying $T$ the dynamical properties of the system change from
regular to chaotic. The first Lyapunov exponent reaches its maximum for
$T$ comparable to the typical time of the unperturbed flow.

The constraint $V_{top}^2+V_{bot}^2=1$ can be looked
at as a limited energy supply to the system.
\vspace{0.5cm}

{\centerline{\bf Numerical results}}
\vspace{0.6truecm}

We show here the numerical results for inert transport under the
stirring of the above flows.
In order to simulate more realistic dispersion processes, we integrate
Eq.(\ref{eq:La}), including the effect of a nonzero molecular
diffusivity $D_0$.

\begin{figure}[htbp]
\centerline{\includegraphics[width=5cm]{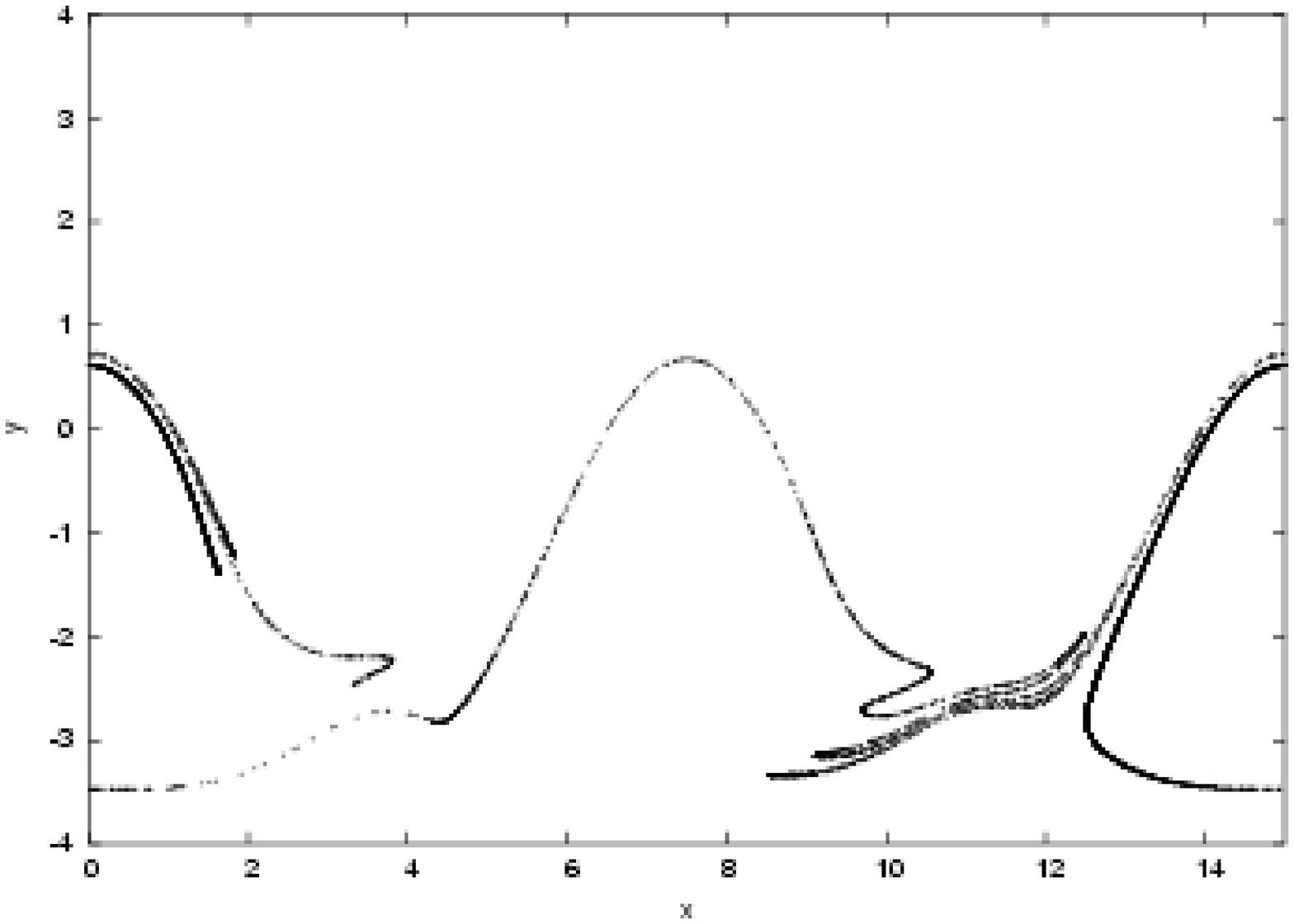} 
            \hspace{0.5cm}
            \includegraphics[width=5cm]{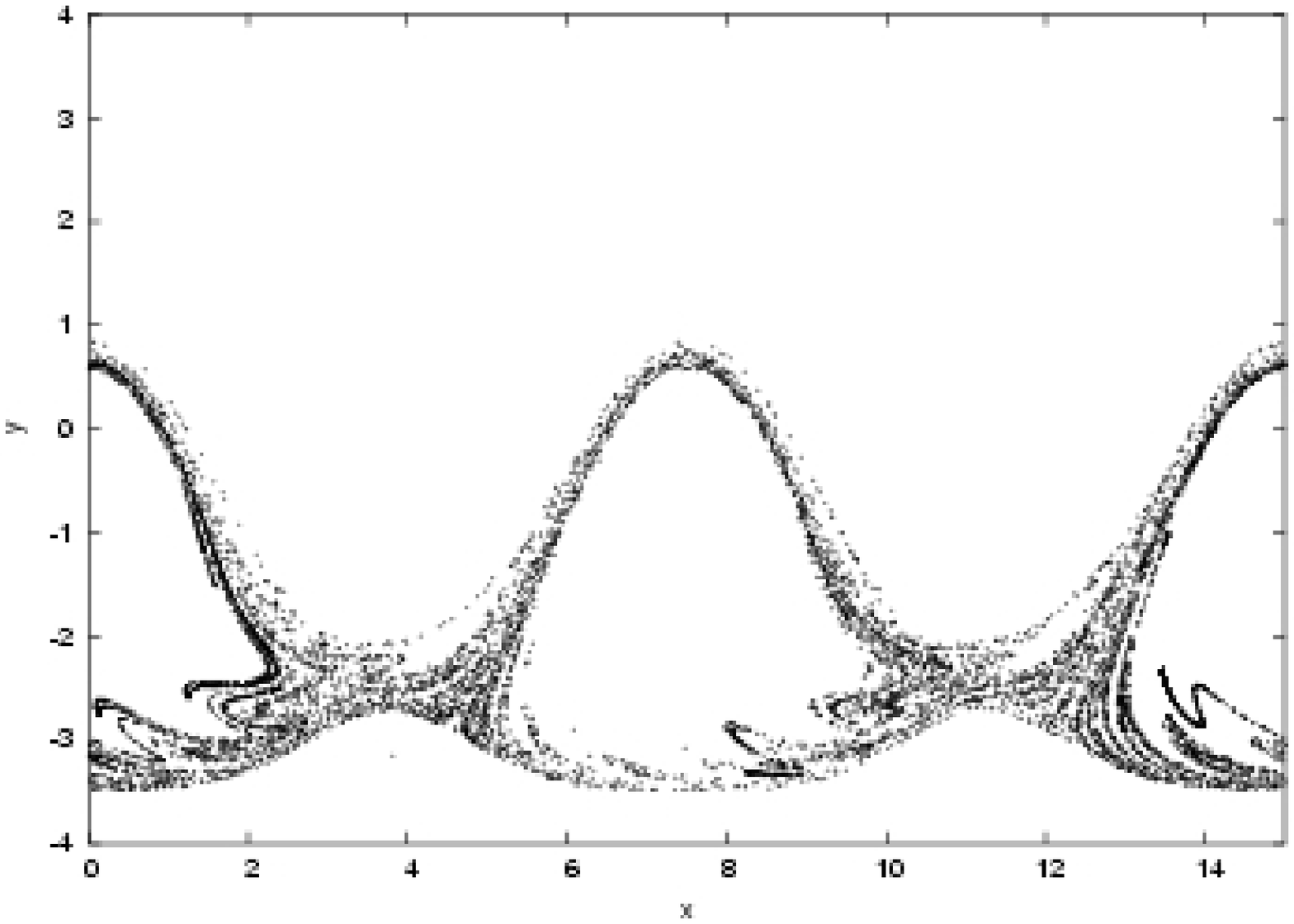}
            \hspace{0.5cm}
            \includegraphics[width=5cm]{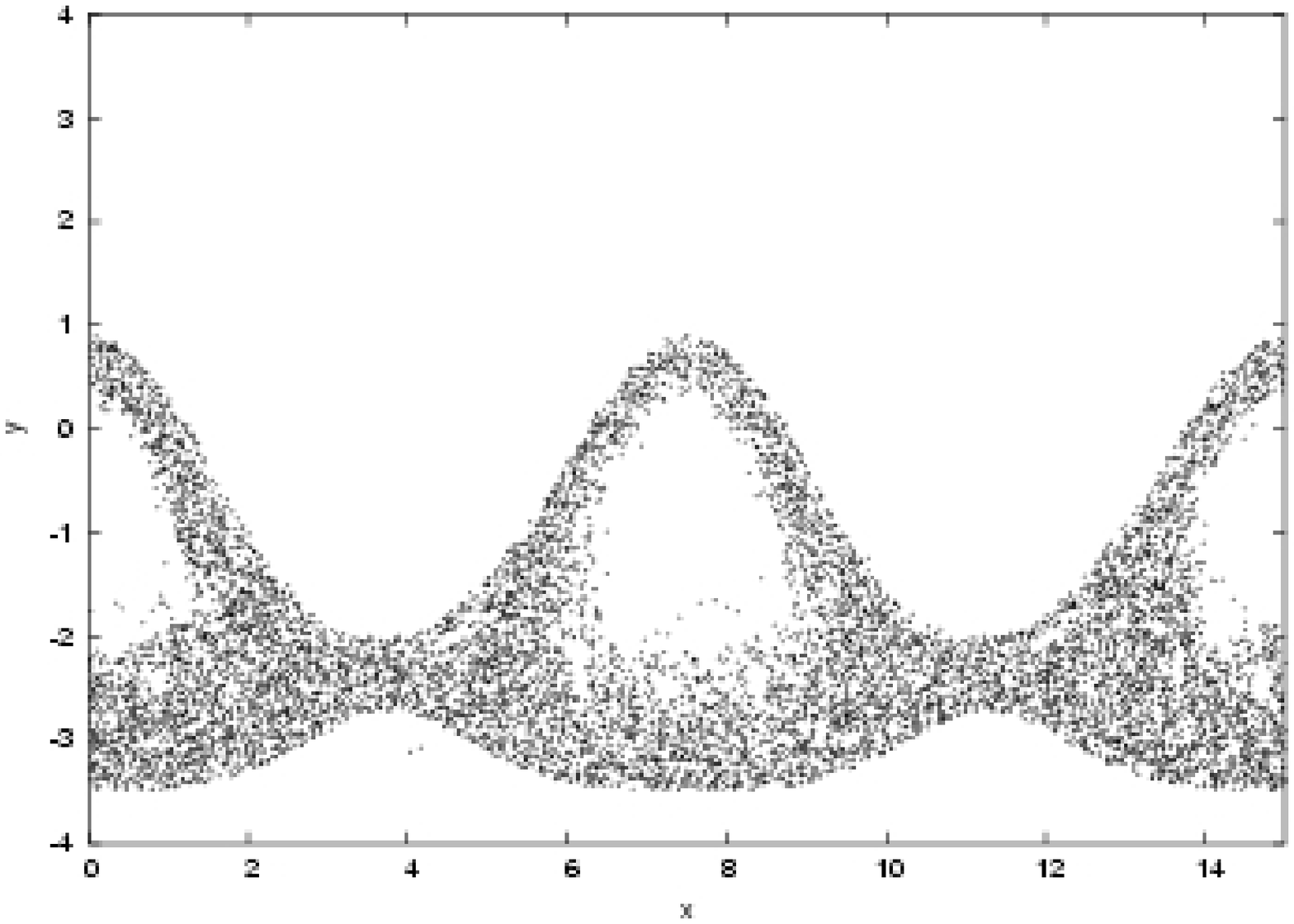}}

\centerline{\includegraphics[width=5cm]{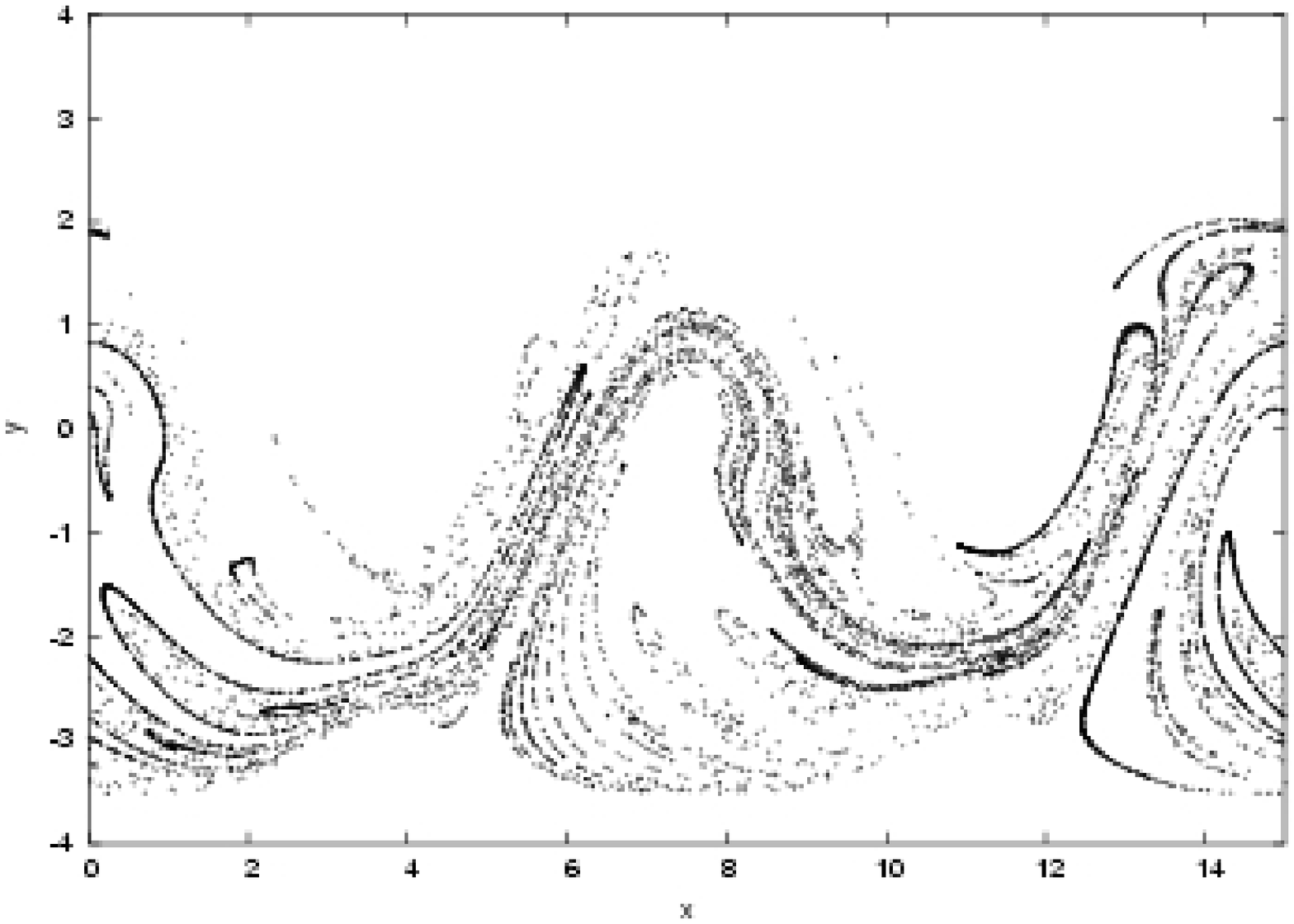}
            \hspace{0.5cm}
            \includegraphics[width=5cm]{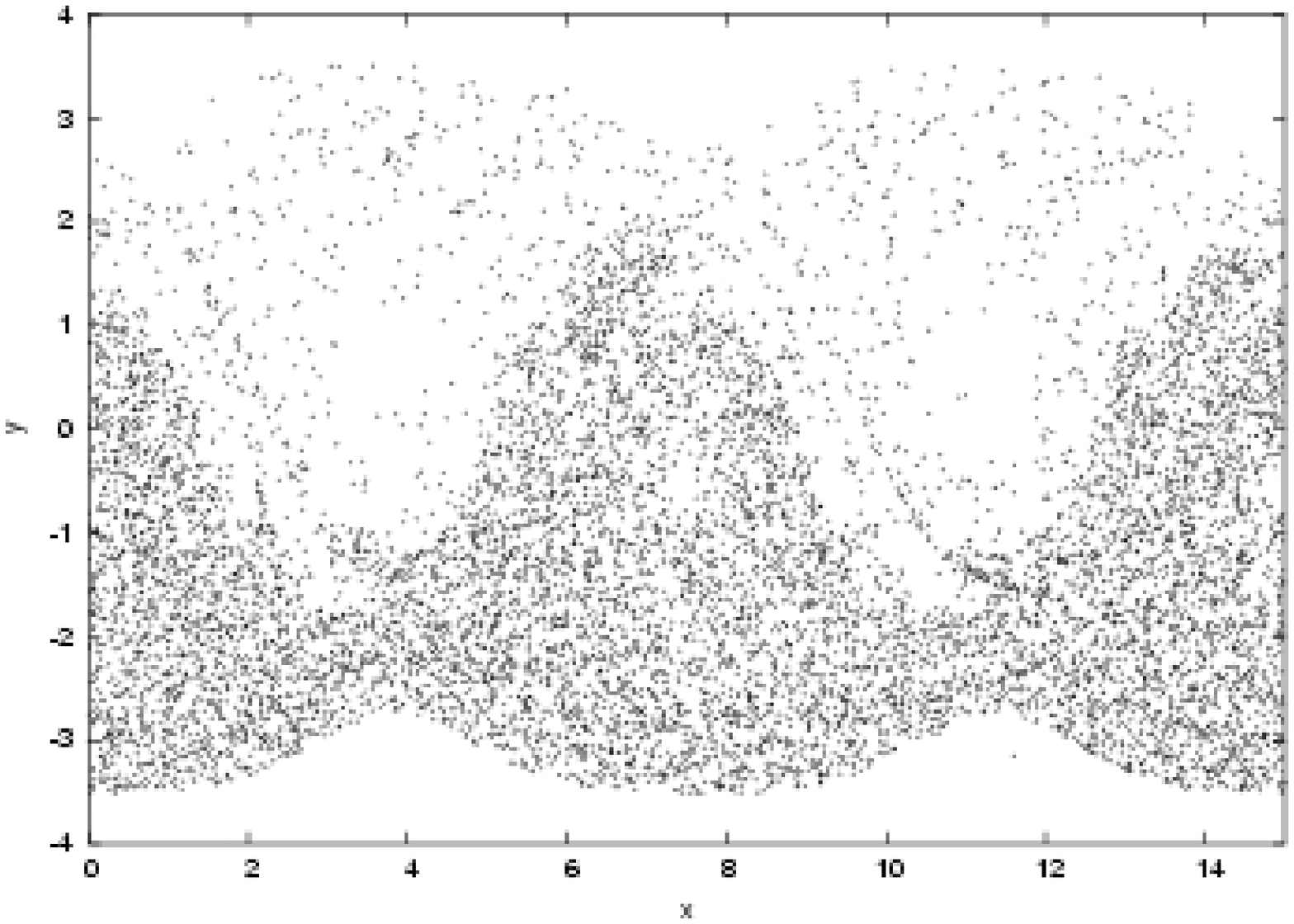} 
            \hspace{0.5cm}
            \includegraphics[width=5cm]{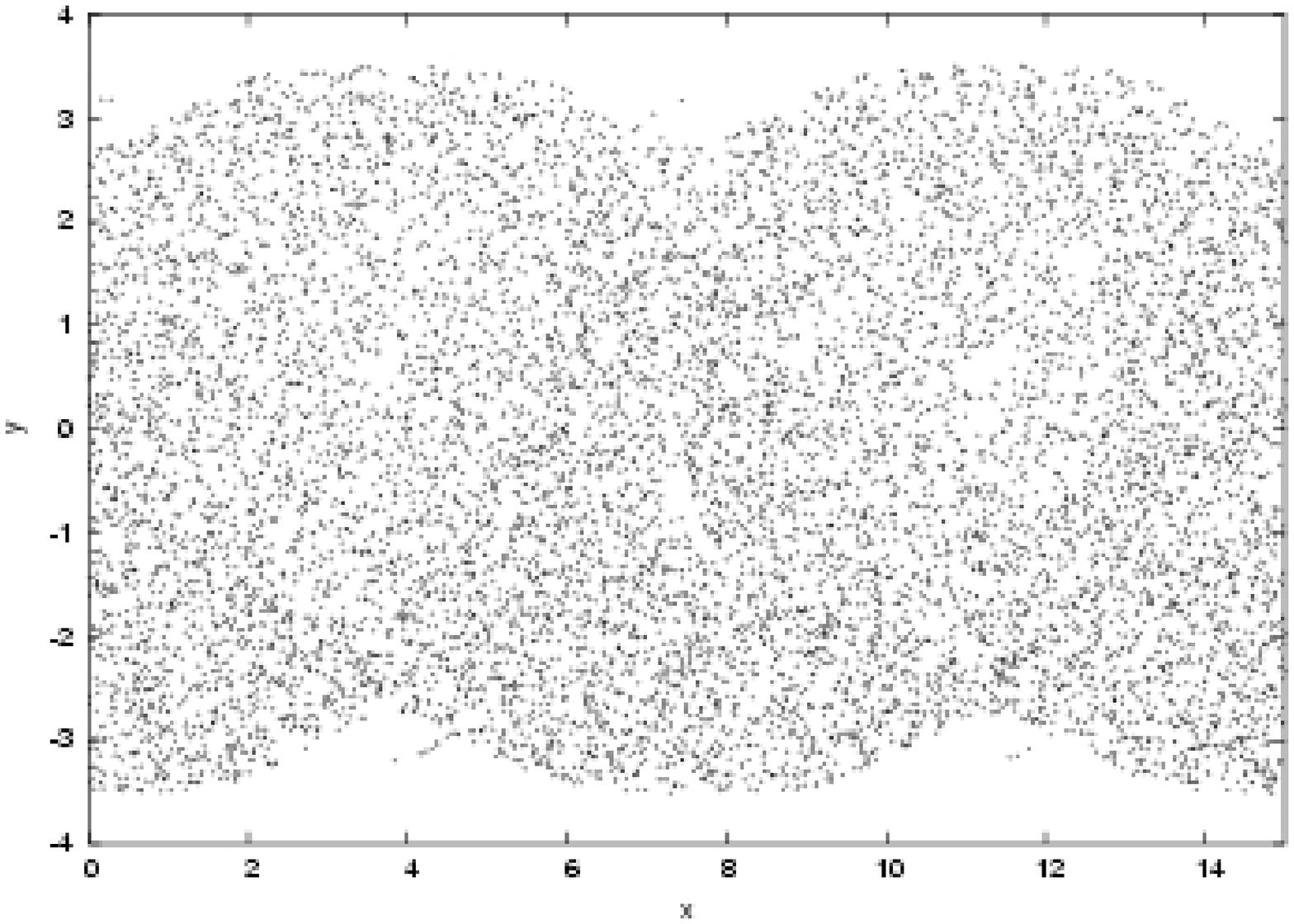}}

\caption{Meandering jet: dispersion of $10000$ particles at times
        (from the left to the right): $t=15,30,200$, in units
        of $T=2\pi/\omega=2\pi/0.625$ (perturbation's period);
        top: $\epsilon=0.03$ (local chaos regime),
        bottom: $\epsilon=0.24$ (large-scale chaos regime).}
\label{fig:mjet-disp}
\end{figure}

\begin{figure}[htbp]
\centerline{\includegraphics[width=7.5cm]{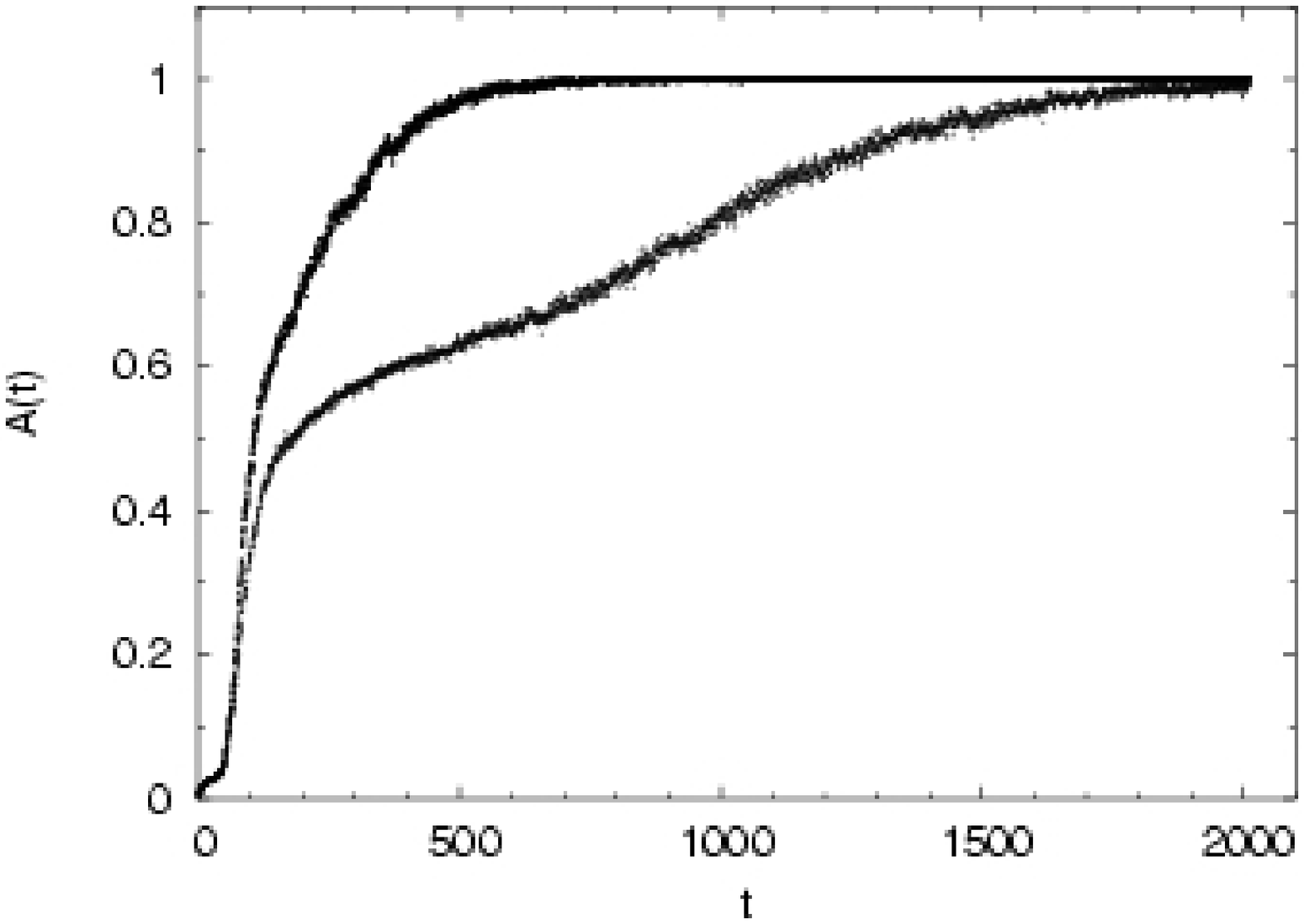} 
            \hspace{0.5cm}
            \includegraphics[width=7.5cm]{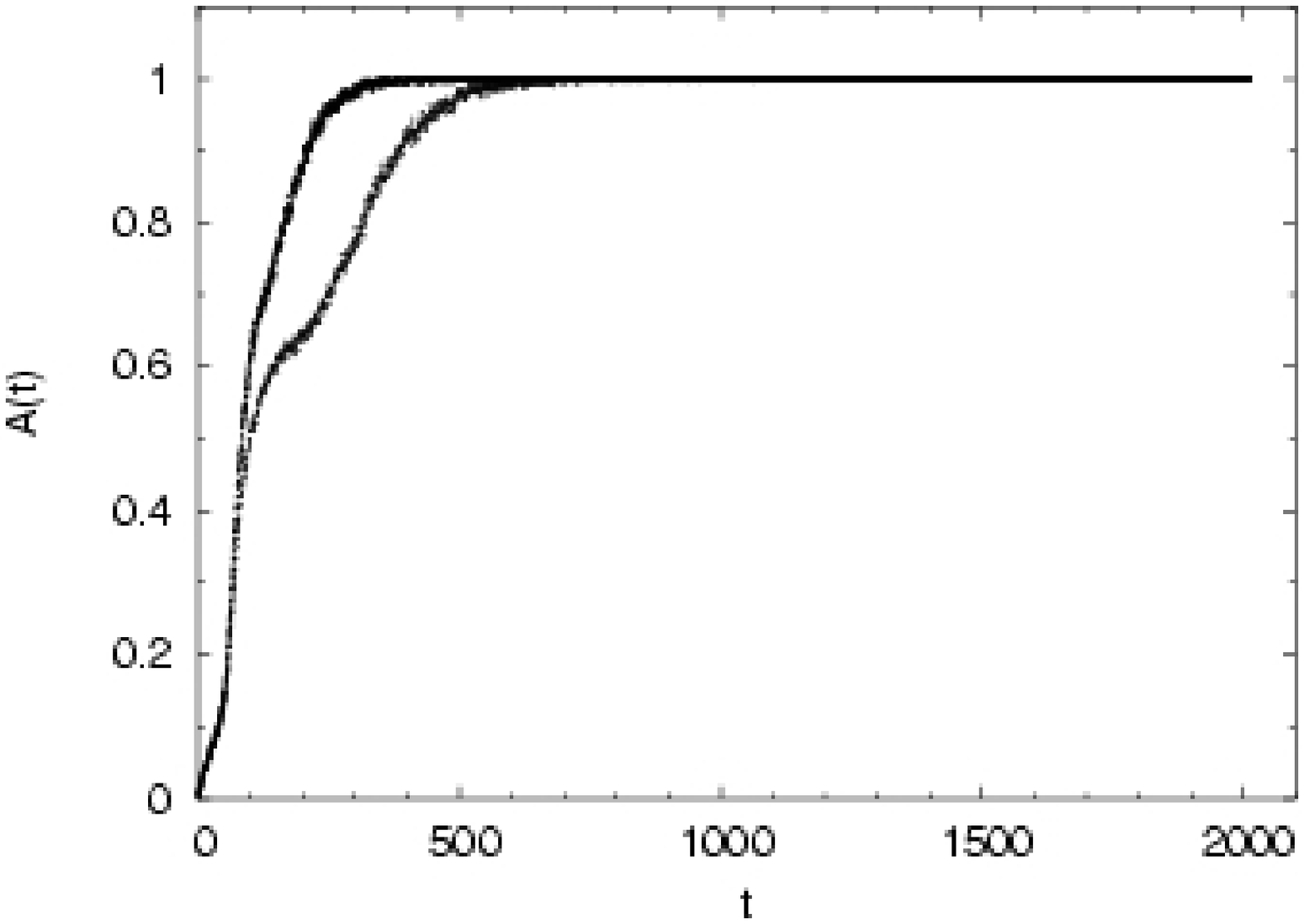}}
\caption{Meandering jet: occupied area vs $t$, on the left $D_0=0.001$, on
the right $D_0=0.004$.  Top curve shows the case with overlap 
of resonances $(\epsilon=0.24,\omega=0.625)$
(large scale chaos regime), 
bottom curve shows the case with nonoverlap of resonances
$(\epsilon=0.03, \omega=0.625)$
(local chaos regime).}
\label{fig:oa}
\end{figure}

The study is carried out following the
time evolution of a cloud of ${\mathcal N} \gg 1$ test particles, initially 
located in a small square of linear size $\delta_\low{0} \ll L$ 
(see Fig.~\ref{fig:mjet-disp}). 
We show the behavior of the two systems in the local chaos and large-scale 
chaos regime. We present mainly the result for the meandering jet;
similar behaviors have been observed also in the Stokes flow. 
The role of the two different dynamical regimes is clearly seen in
Fig.~\ref{fig:oa}, where the fraction of occupied area is shown. 
The curves related to local chaos are always lower than the ones
related to large-scale chaos, and the saturation times needed to invade 
the whole domain's area are significantly different in the two cases.

For small values of the molecular diffusion coefficient, in the non
overlap situation for system (\ref{eq:mjet-sf}) it is possible to see a
slowing down of the process, around half of the total area, due to the
diffusive crossing of the jet. Let us incidentally note that, for $D_0=0$,
no crossing of the jet would be possible for $\epsilon<\epsilon_c$. This
can be seen in Fig.~\ref{fig:mjet-disp}, where the dispersion of a
cloud of test particles is plotted for the two dynamical regimes.
Anyway, the presence of the noisy term (i.e., $D_\low{0} > 0$)
does not change the scenario if $D_\low{0}$ is small. 
See Fig.~\ref{fig:oa}.

The differences in the saturation times progressively diminish for
growing values of $D_0$ (left and right part of Fig.~\ref{fig:oa}).
Indeed, molecular diffusivity helps the dispersion process, acting
itself as a mixing mechanism, though stochastic and not
deterministic. For sufficiently large $D_0$ $[{\mathcal O} (10^{-2})$
- not shown here$]$, no difference is observable, due to the more
relevant weight of the stochastic term with respect to the
deterministic one in the Langevin equation (\ref{eq:La}). 

\begin{figure}[htbp]
\centerline{\includegraphics[height=6cm]{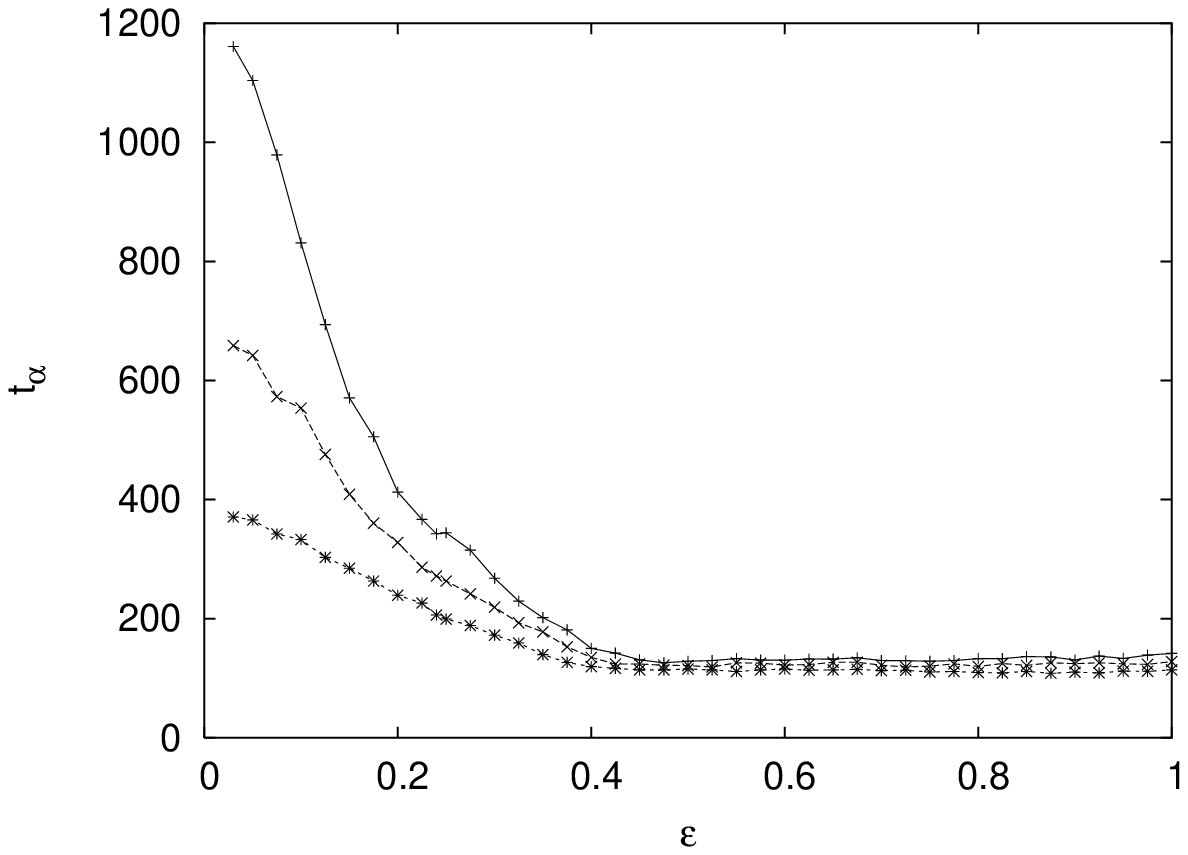}
            \hspace{0.5cm}
\includegraphics[height=6.6cm]{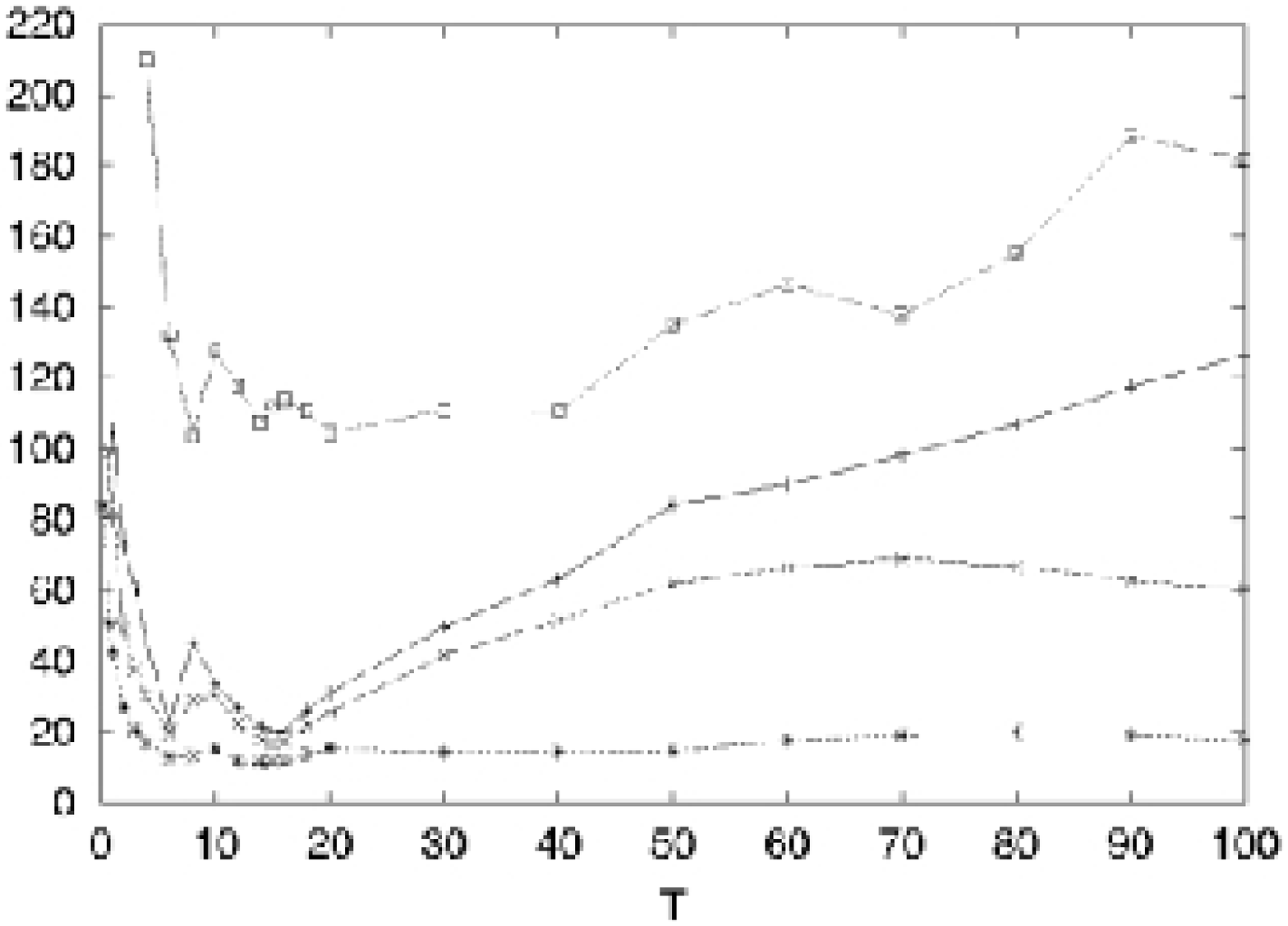}}
\caption{On the left, mixing times in the meandering jet: 
$t_{\alpha}$ vs $\epsilon$ at $\omega=0.625$.  From the top to the
bottom: $D_\low{0}=0.001,\:0.002,\:0.004$.  On the right, Stokes flow:
mixing times and inverse of Kolmogorov-Sinai entropy ($c/h_{KS}$)
vs $T$.  The time $c/h_{KS}$ (on top) is computed for $D_0 = 0$,
mixing time curve are, from top to bottom $D_\low{0}=0.0005,
\:0.001, \:0.004$.}
\label{fig:mt}
\end{figure}

The behaviors of $t_\alpha$ as function of $t$ (Fig.~\ref{fig:mt})
allow a more extensive analysis.
With the same set of parameters as before,
a wider scan in the values of chaos control parameters
($\epsilon$ for meandering jet, $T$ for Stokes flow) has been carried
out for $\alpha=0.9$ and various values of $D_0$.
As expected, these times slowly vary with the molecular diffusion
coefficient in a strongly chaotic dynamical regime generated only by
the flow. The dependence on $D_0$ becomes stronger when the dynamics
only due to $\bu$ is almost regular.  In fact, our results show the
great relevance of the details of the velocity field on mixing
efficiency (see Fig. \ref{fig:mt}).

Moreover, we show mixing times for the Stokes flow
together with an entropy graph. We plotted the quantity $c/h_{KS}$
versus $T$: $1/h_{KS}$ has dimensions of time and $c$ is a 
dimensionless parameter.
Let us note that both $t_\alpha$ and $1/h_{KS}$ have minima
for those values of $T$ which give the most chaotic
dynamics.

We also remark that numerical values of our observables can depend on
initial positions of test particles, but qualitative behaviors are
general.

Let us observe that in the purely diffusive case ($\bu=0$)
$t_{\alpha}$ is nothing but the "bare" diffusive time needed to invade
the whole domain.  In that case we recovered the inverse
proportionality relation $t_{\alpha} \sim L^2/D_0$.
\section{Reactive Case}
\label{sec:3}

Now we deal with the complete equation~(\ref{eq:ARD}) studying an ARD
system confined in a closed domain.  As an initial condition we consider
a small quantity of active material, i.e., $\theta=1$ in a small
region of $\Omega$ with linear size $\delta_\low{0}$, 
elsewhere $\theta=0$. We numerically compute
the time needed for a given percentage of the total area to be filled
by the reaction (called, in the following, the reacting or burning
time).  A natural and important question is how the burning time
depends on the transport properties of the flow.

The velocity fields are the already-presented meandering jet 
and Stokes flow. Our main result,
obtained in both flows, is that the burning time is not 
strictly related to the transport properties of the flow.

As for the case of inert transport, the principal observable under 
investigation is the time needed for a given percentage of the total 
area to be burnt. We define 
\begin{equation}
S(t) =\frac{1}{|\Omega|} \int_{\Omega} 
{\mathrm d}x {\mathrm d}y \theta (x,y,t),
\label{eq:burnarea}
\end{equation}
as the percentage of area burnt at time $t$, where $|\Omega|$ is the
area of the domain $\Omega$. In our case, we choose an appropriate
localized initial condition such that the initial burnt material is
$S(0) = 0.005$.  The reacting or
burning time $t_\alpha$ is defined as the time needed for the
percentage $\alpha$ of the total area of the recipient to be burnt,
i.e.,
\begin{equation}
S(t_\alpha)=\alpha \,\,.
\label{eq:burneff}
\end{equation}
\begin{figure}[htbp]
\centerline{\includegraphics[width=5cm]{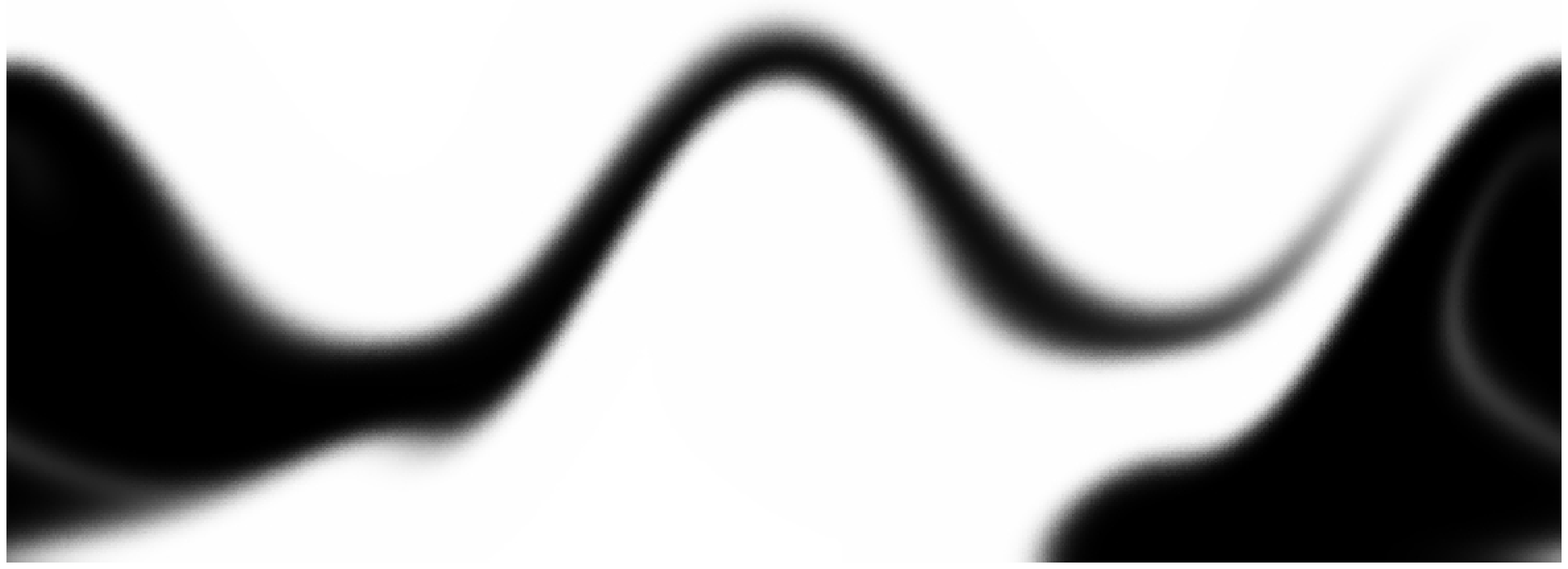} 
            \hspace{0.5cm}
            \includegraphics[width=5cm]{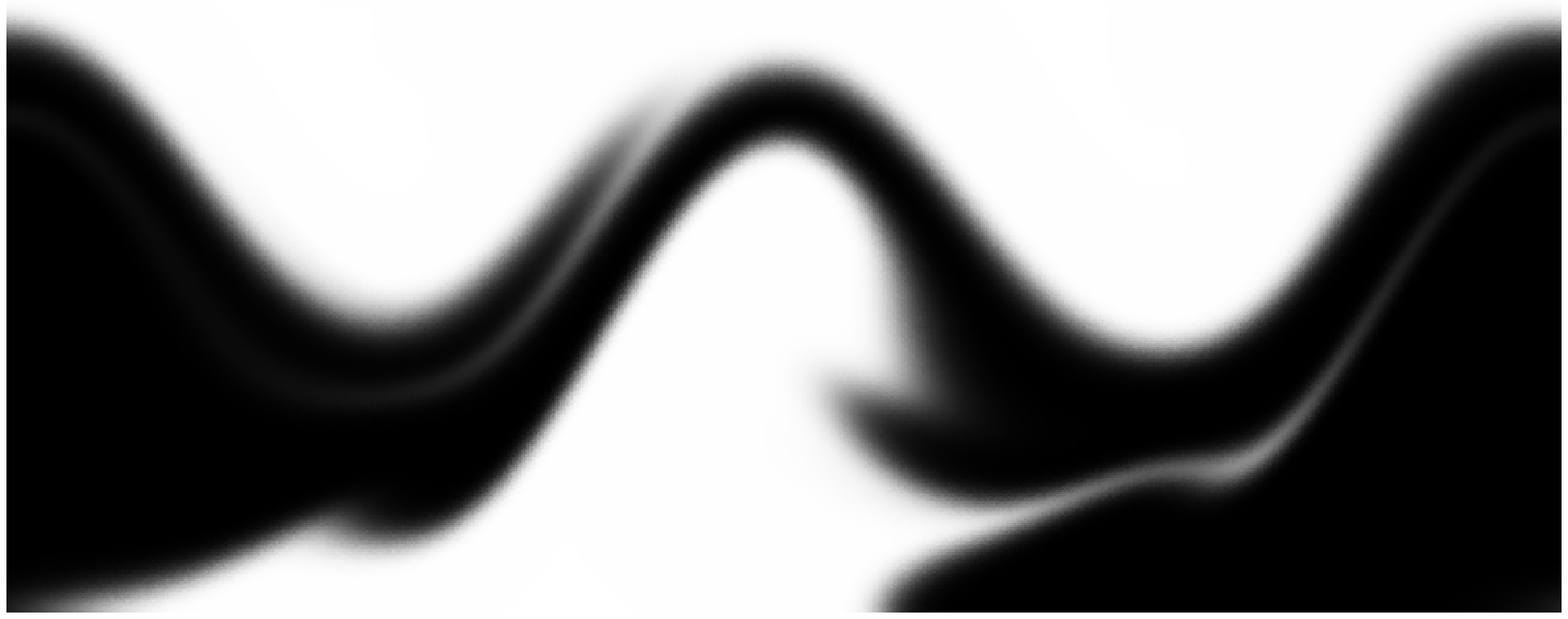}
            \hspace{0.5cm}
            \includegraphics[width=5cm]{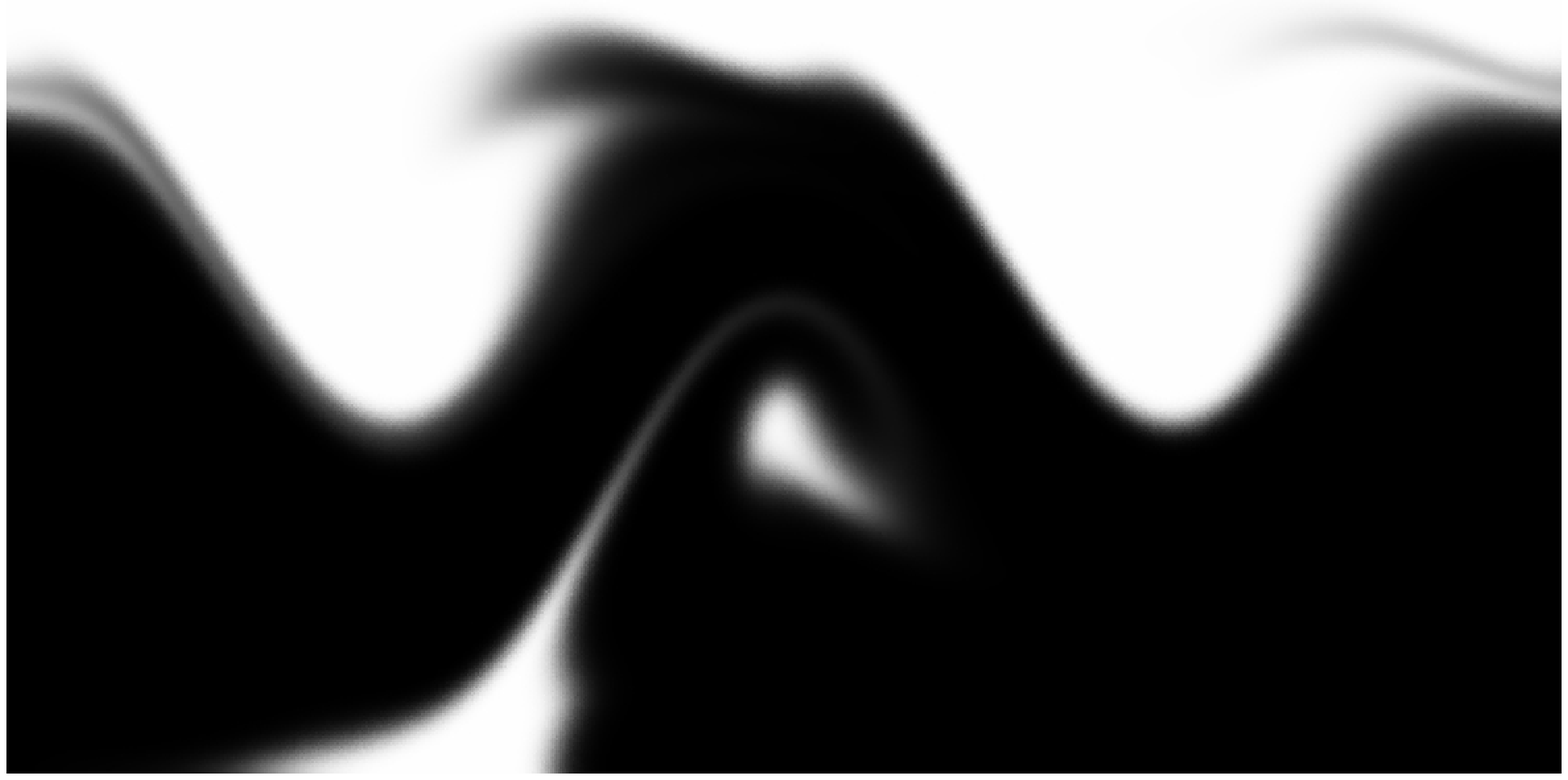}}

\centerline{\includegraphics[width=5cm]{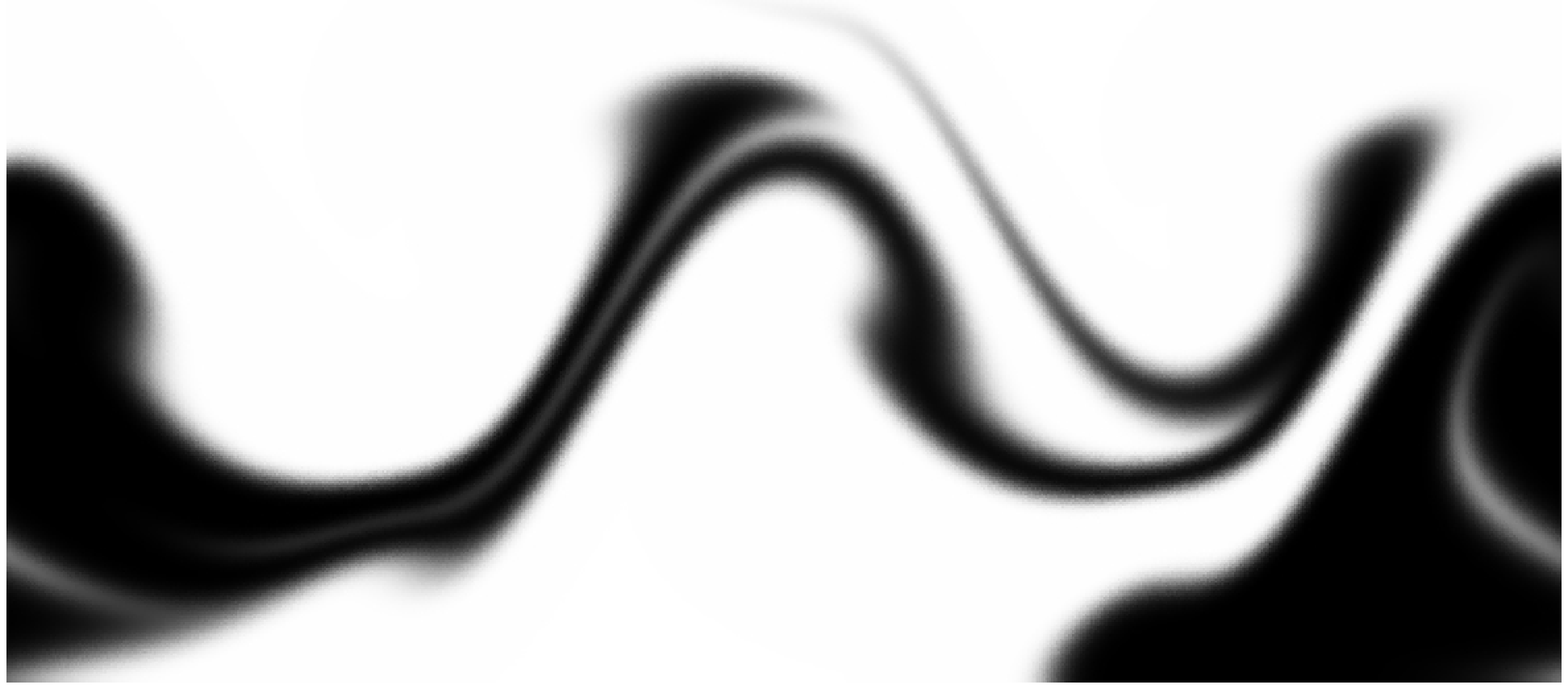}
            \hspace{0.5cm}
            \includegraphics[width=5cm]{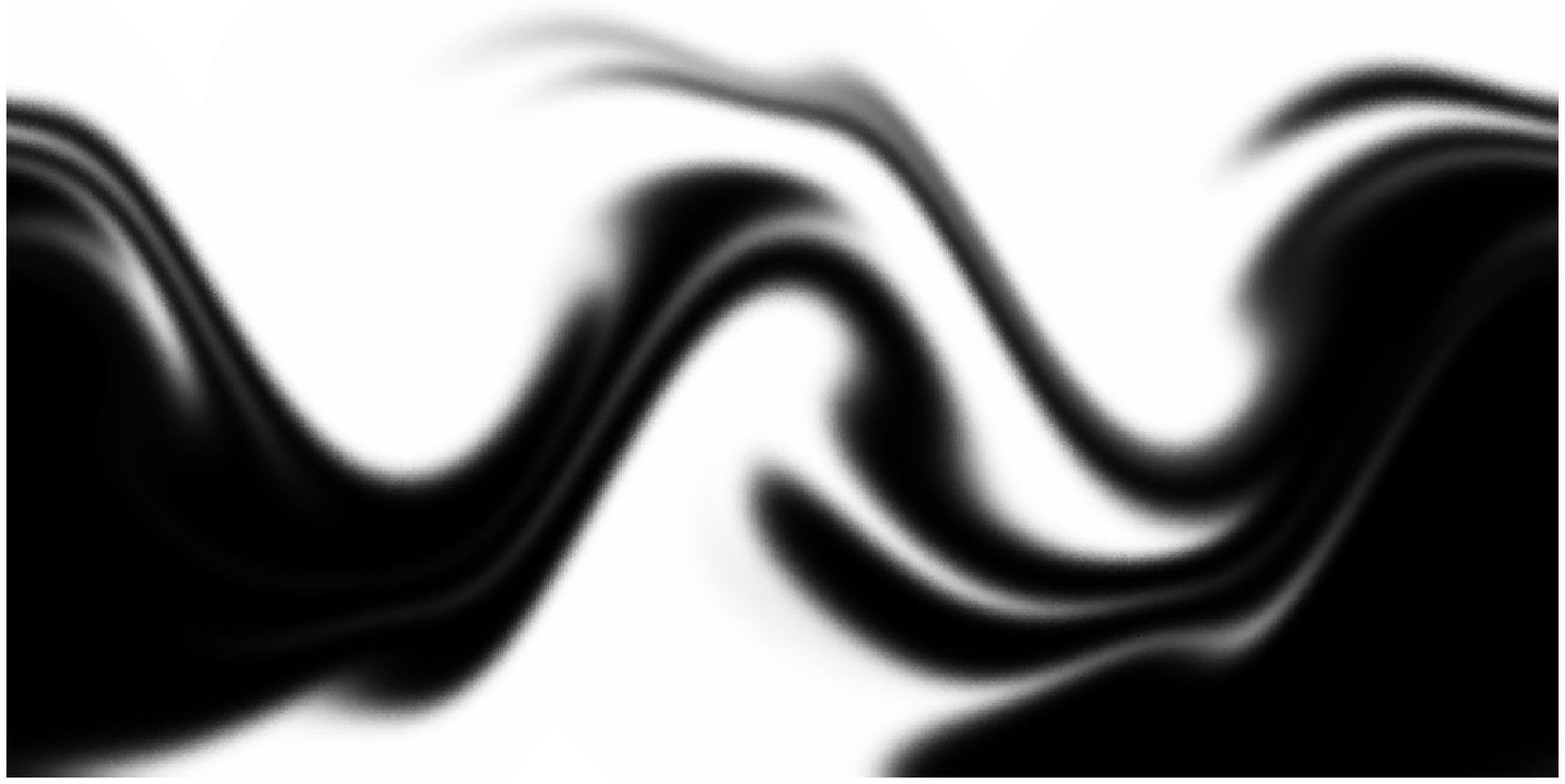} 
            \hspace{0.5cm}
            \includegraphics[width=5cm]{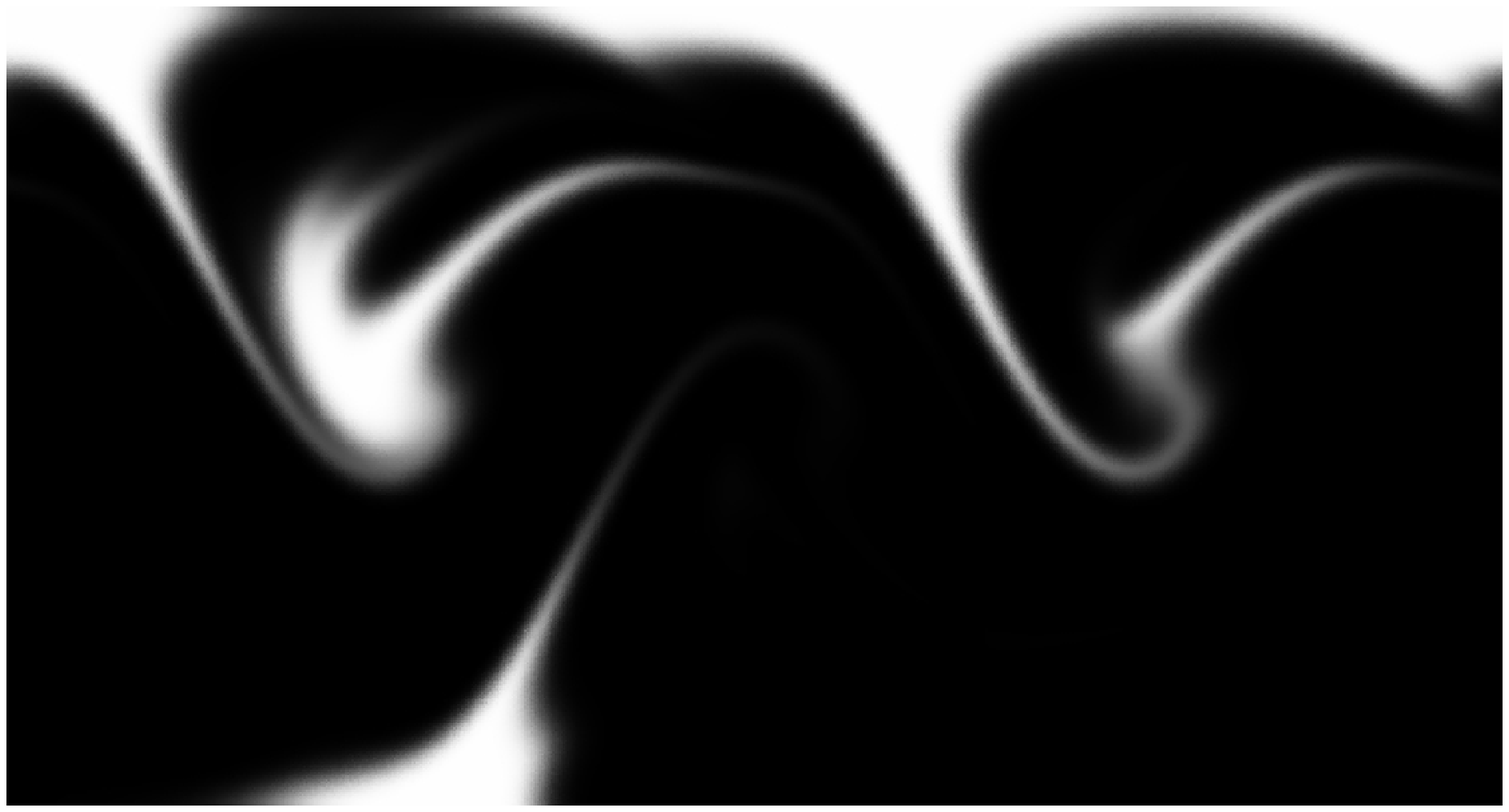}}
\vspace{0.5truecm}

\caption{Meandering jet ($2\ell$): Snapshots of the field 
$\theta(x, y, t)$ at times (from the left to the right): $t=4,5,7$,
in units of $T=2\pi/\omega=2\pi/0.625$ (perturbation's
period); top row: $\epsilon=0.03$ (local chaos regime), bottom
row $\epsilon=0.24$ (large-scale chaos regime); $D_\low{0}=0.001$.
Black corresponds to $\theta = 1$; white indicates $\theta = 0$.}
\label{fig:btheta}
\end{figure}
\vspace{0.5truecm}

To numerically integrate Eq.~(\ref{eq:ARD}) we followed
a pseudo-Lagrangian approach.
This algorithm uses a path integral formulation for $\theta(x,y,t)$: 
the field evolution is computed using the Lagrangian
propagator plus a Monte Carlo integration for the diffusive term; 
then, the reaction propagator accounts for the reacting term
(for details see Ref.~\cite{abe01}). 

We impose a rigid wall condition in the
boundaries, in order to avoid any fluid particle leaving the
container, which could happen due to the noise term added to the
velocity field in the Lagrangian approach.

Figure~\ref{fig:btheta} shows some snapshots of the
concentration field in the meandering jet flow for two different values
of the control parameter. Comparing this figure with the analogous
Fig.~\ref{fig:mjet-disp} in the inert case (using the same
set of parameter), it is possible to have a clear insight
of the different behavior of the system when reaction is present.\\
\begin{figure}[htbp]
\includegraphics[width=7cm]{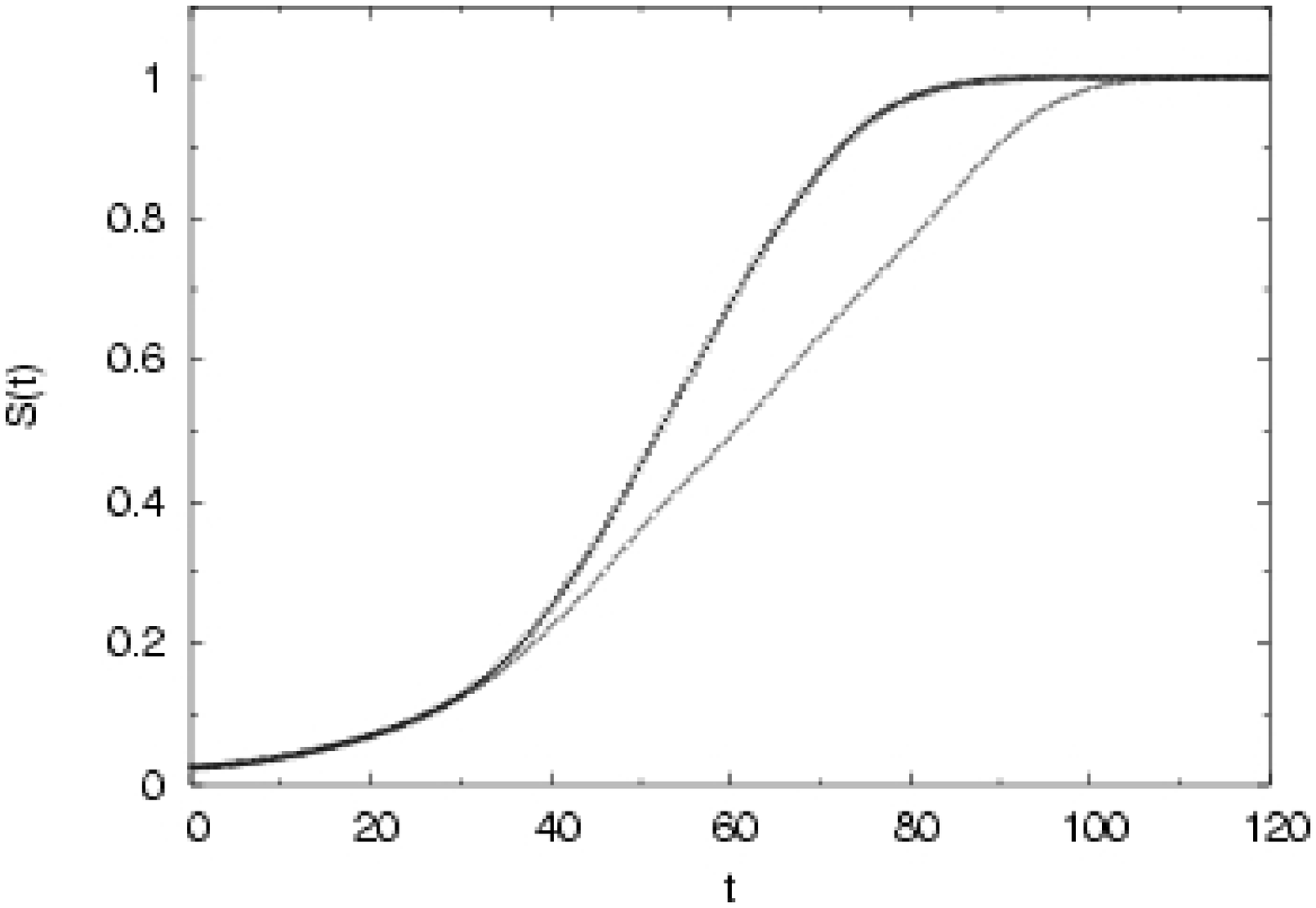}
\hspace{1.0cm}
\includegraphics[width=7cm]{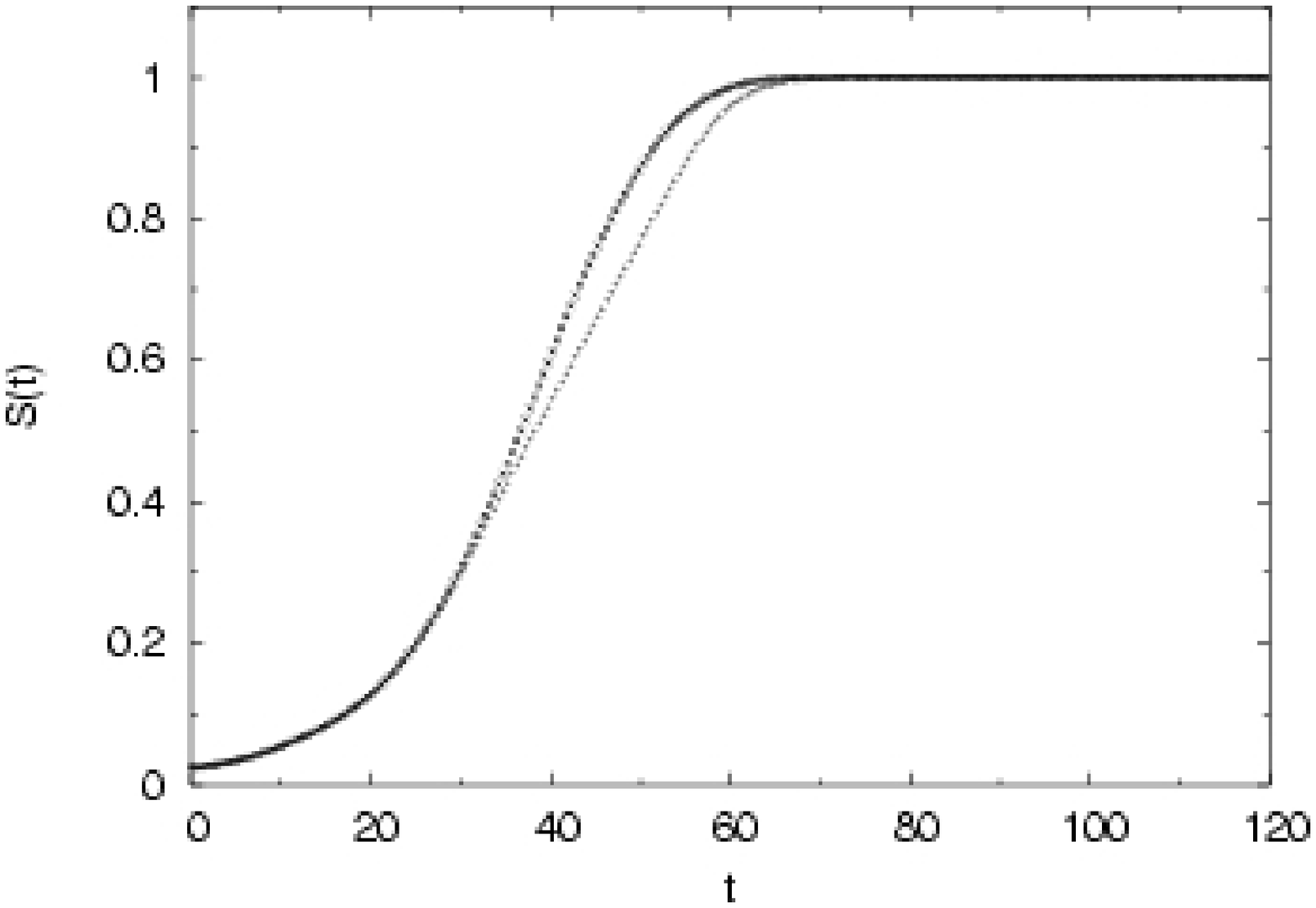}
\vspace{0.3cm}

\caption{Meandering jet. Burnt area vs $t$, on the left
$D_\low{0}=0.001$, on the right $D_\low{0}=0.004$,
$\tau=2$, top curve $(\epsilon=0.24, \omega=0.625)$ 
(large-scale chaos regime),
bottom curve $(\epsilon=0.03, \omega=0.625)$
(local chaos regime).}
\label{fig:reaco}
\end{figure}

\begin{figure}[htbp]
\includegraphics[width=7cm]{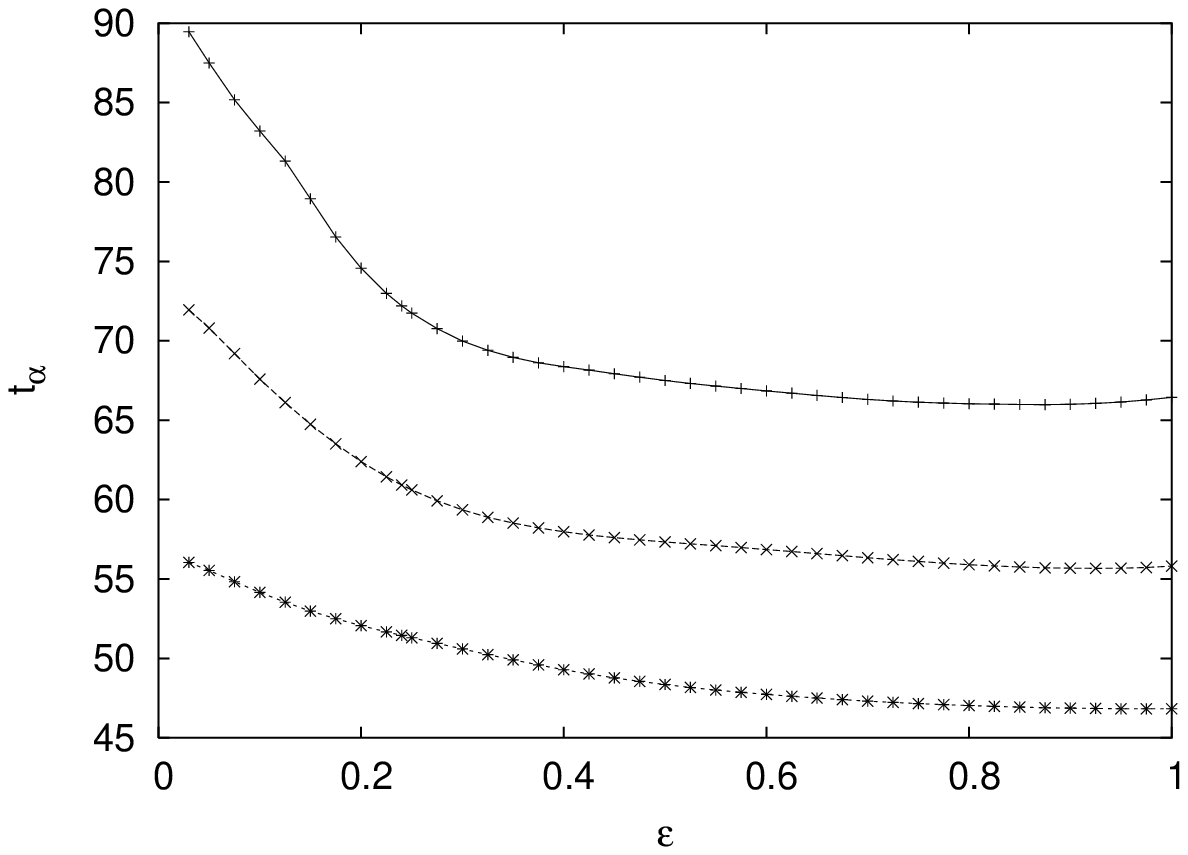}
\hspace{1.0cm}
\includegraphics[height=5.4cm]{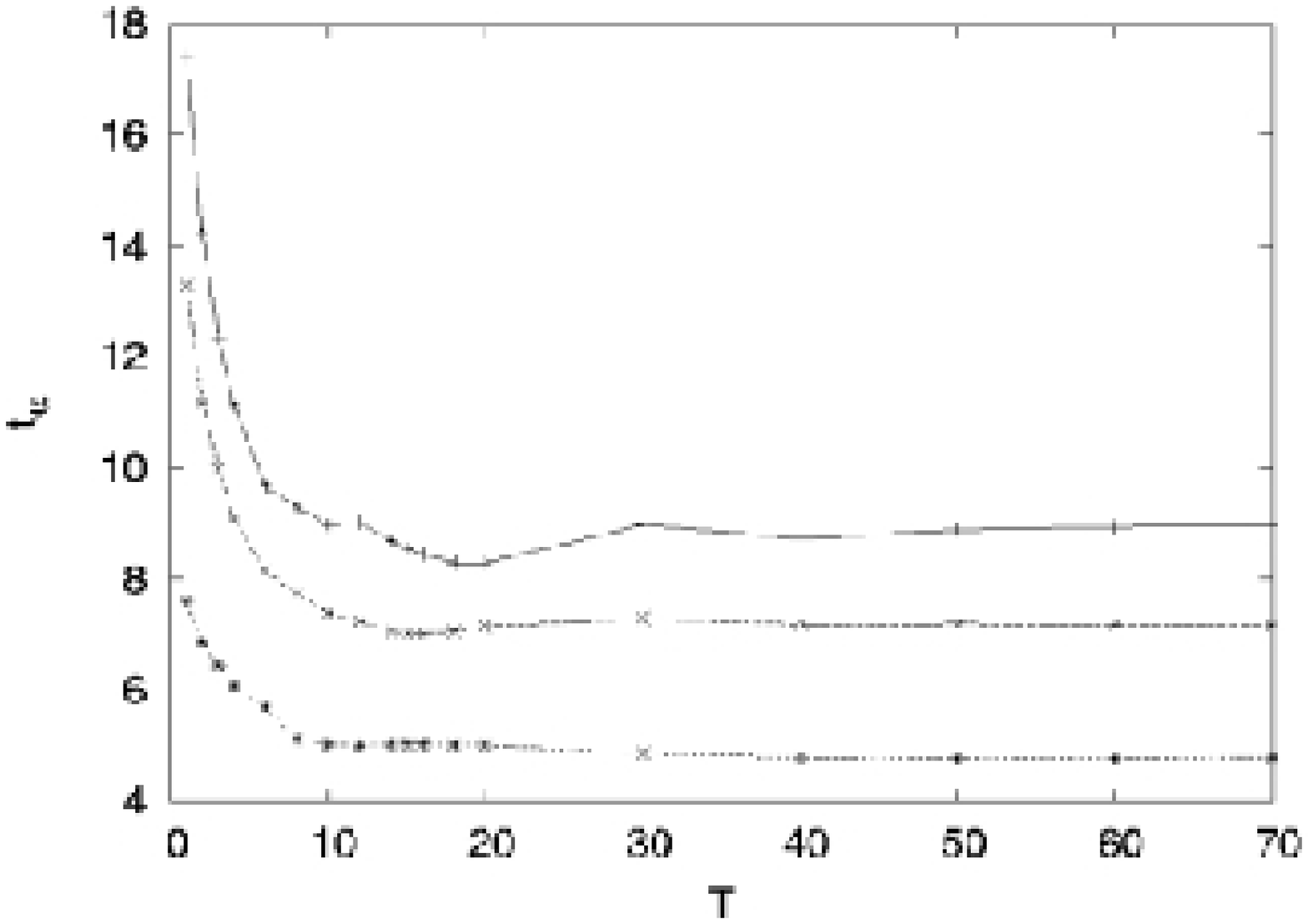}
\vspace{0.3truecm}

\caption{Burning times vs $t$. On the left meandering jet. 
$t_{\alpha}\:(\alpha=0.9)$ vs $\epsilon$, $\omega=0.625$
from the top to the bottom: $D_\low{0}=0.001,\:0.002,\:0.004$.
On the right, Stokes flow. From top to bottom 
$D_0= \:0.0005, \:0.001,\:0.004$. In both cases the time scale 
of the chemistry is $\tau=2$.}
\label{fig:be}
\end{figure}

Figure~\ref{fig:reaco}, which shows the burnt area as a function of time 
in the case of local and large-scale chaos, has to be compared to the
analogous Fig.~\ref{fig:oa} regarding the occupied area.  It is apparent
that, passing from local to large-scale chaotic dynamics, while the
mixing efficiency changes greatly, the burning efficiency varies
only slightly.  This is a first evidence of the regularization
properties of the reaction term. A further confirmation of such a
feature comes from Fig.~\ref{fig:be}, where the burning efficiency is shown 
for different control parameters of the flows. In fact,
it is possible to see that, at varying the control parameters, 
the burning efficiency changes only slightly.
Such a behavior is very different from that observed for
the mixing efficiency (see Fig.~\ref{fig:mt}).
Let us note that for the Stokes flow different values of the control
parameter $T$ give similar mixing time, but different burning time 
(compare the right part of Figs.~\ref{fig:mt} and ~\ref{fig:be}).
Therefore, the burning efficiency 
is not so strictly related to the mixing properties of the flow.
From Fig.~\ref{fig:be} the presence of a
plateau (and a consequent lower bound for the burning time) appears in
the burning efficiency.  As shown in Ref.~\cite{lop02} this plateau depends
mainly on the reaction characteristic time.  \\

\begin{figure}[htbp]
\centerline{\includegraphics[width=8cm]{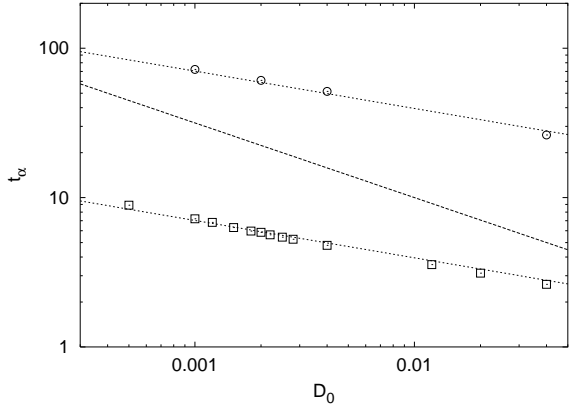}}
\vspace{0.3truecm}

\caption{Burning times vs $D_\low{0}$ for the meandering jet ($\circ$)
and the Stokes flow ($\Box$) in the plateau
region. The dotted lines show the behavior $t_\alpha \sim D_\low{0}^{-1/4}$;
the dashed line shows the behavior $t_\alpha \sim D_\low{0}^{-1/2}$.}
\label{fig:taD0}
\end{figure}

The above results confirm the subtle and intriguing combined
effect due to Lagrangian chaos, diffusion, and reaction.
This issue is important to many different fields including
the classical limit of quantum mechanics~\cite{quantum}.
In Ref.~\cite{Cencini2003} it is shown that, at variance with the inert 
transport~\cite{Castiglione}, for the asymptotic front propagation
properties, the role of the Lagrangian chaos is marginal
if diffusion and reaction are present. In this preasymptotic
problem we have that, independently of the details of $\bu$
(in the presence or not of large-scale chaos) the dependence
of $t_\alpha$ on $D_\low{0}$ is rather weak.
In Fig.~\ref{fig:taD0} we show $t_\alpha$ at varying $D_0$
in the plateau region for the Stokes flow
and the meandering Jet.
We have fair evidence that 
$t_\alpha \sim D_\low{0}^{-1/4}$, which is rather different
from the result in absence of $\bu$, i.e., 
$t_\alpha \sim L/v_\low{f} \sim D_\low{0}^{-1/2}$.
In other words, in the reacting case, the combined
effect of the advection reaction and diffusion allows
an efficient non trivial ``crossing of the dynamical
barriers''. 

\begin{figure}[htbp]
\centerline{\includegraphics[width=7cm]{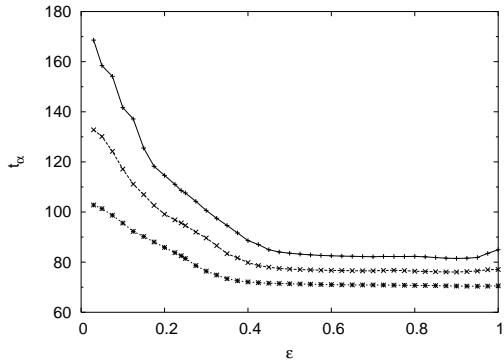}}
\vspace{0.3truecm}

\caption{Burning times for the meandering jet
with an Arrhenius reaction term,
$t_{\alpha}\:(\alpha=0.9)$ vs $\epsilon$, $\omega=0.625$
from the top to the bottom: $D_\low{0}=0.001,\:0.002,\:0.004$.
The Arrhenius parameters are $\tau=2$ and $\theta_c = 0.5$.}
\label{fig:bear}
\end{figure}

It is rather natural to wonder about the generality of the above
results if one changes the term $f(\theta)$ using a non FKPP-type reaction
term. The interesting case is when $f'(0) = 0$, for which the lower
bound for the propagation velocity is 0. In this case, situations  
exist in which the presence of a velocity field
can suppress front propagation~\cite{quenching}.
However, if the reaction takes place in a closed domain,
the reaction term is not pathological [we use the Arrhenius term
$f(\theta) = (1-\theta)e^{-\theta_c/\theta}$] and the
initial size of the active spot $\delta_\low{0}$ is not too small,
the qualitative scenario showed above does not change.
Figure~\ref{fig:bear} gives clear evidence of this.
The behavior $t_\alpha \sim D_\low{0}^{-1/4}$ in the plateau region 
(see Fig.~\ref{fig:taD0}) is confirmed also in the Arrhenius case.

\section{Conclusion}
\label{sec:4}

We have performed a numerical study of advection-reaction-diffusion
systems confined in a closed domain and stirred by two different
laminar velocities. Both the velocity fields can generate a regular
or a chaotic Lagrangian dynamics, at varying control parameters.
For the mixing properties of inert particles we observed
that, when the dynamics is strongly chaotic, mixing times 
weakly depend on molecular diffusion; this feature becomes 
much more notable when the velocity field {\bf u} is not 
strong enough to avoid the creation of recirculation regions.

Then, switching on the reaction term, we analyze the burning 
time of a reactive scalar in the same flows. 
The principal result of our study is that, while the mixing 
properties of the flows can change very much
with varying dynamical properties,
on the contrary the burning efficiency does not vary so much. 
We have also shown 
cases exist in which the burning efficiency
is not strictly related to the mixing properties of the flows.
Moreover, all the previous results are quite independent from
the shape of the reaction term, $f(\theta)$.
\\

We thank Massimo Cencini for a careful reading of the manuscript.

\end{document}